\documentclass[fleqn,usenatbib]{mnras}

\usepackage{multirow}

\usepackage{newtxtext,newtxmath}
\usepackage[T1]{fontenc}

\DeclareRobustCommand{\VAN}[3]{#2}
\let\VANthebibliography\thebibliography
\def\thebibliography{\DeclareRobustCommand{\VAN}[3]{##3}\VANthebibliography}

\usepackage{graphicx}	% Including figure files
\usepackage{physics, amsmath}	% Advanced maths commands
\usepackage{hyperref}

\title[QSOs eBOSS BAO measurements with voids]{Void BAO measurements on quasars from eBOSS}

\author[A. Tamone et al.]{
\parbox{\textwidth}{
Am\'elie Tamone$^{1,}$\thanks{E-mail: \texttt{\href{mailto:amelie.tamone@epfl.ch}{amelie.tamone@epfl.ch}}},
Cheng Zhao$^{1,}$\thanks{E-mail: \texttt{\href{mailto:cheng.zhao@epfl.ch}{cheng.zhao@epfl.ch}}},
Daniel Forero-S\'anchez$^1$,
Andrei Variu$^1$,
Chia-Hsun Chuang$^2$,
Francisco-Shu Kitaura$^{3,4}$,
Jean-Paul Kneib$^{1,5}$,
Charling Tao$^{6,7}$
}
\\
\vspace*{4pt} \\
\small $^{1}$Institute of Physics, Laboratory of Astrophysics, \'Ecole Polytechnique F\'ed\'erale de Lausanne (EPFL), Observatoire de Sauverny, CH-1290 Versoix, Switzerland\vspace*{-2pt} \\
\small $^{2}$Kavli Institute for Particle Astrophysics and Cosmology, Stanford University, 452 Lomita Mall, Stanford, CA 94305, USA\vspace*{-2pt} \\
\small $^{3}$Instituto de Astrofísica de Canarias, s/n, E-38205, La Laguna, Tenerife, Spain\vspace*{-2pt} \\
\small $^{4}$Departamento de Astrof\'isica, Universidad de La Laguna, E-38206, La Laguna, Tenerife, Spain\vspace*{-2pt} \\
\small $^{5}$Aix Marseille Univ, CNRS, CNES, LAM, F13388 Marseille, France\vspace*{-2pt} \\
\small $^{6}$CPPM, Aix-Marseille Universit\'e, CNRS/IN2P3, CPPM UMR 7346, F13288 Marseille, France\vspace*{-2pt} \\
\small $^{7}$Tsinghua Center for Astrophysics, Department of Astronomy, Tsinghua University, Beijing 100084, P.R. China\vspace*{-2pt} \\
}

\date{Accepted XXX. Received YYY; in original form ZZZ}

\pubyear{2022}

\begin{document}
\label{firstpage}
\pagerange{\pageref{firstpage}--\pageref{lastpage}}
\maketitle

\begin{abstract}
We present the clustering of voids based on the quasar (QSO) sample of the extended Baryon Oscillation Spectroscopic Survey Data Release 16 in configuration space. We define voids as overlapping empty circumspheres computed by Delaunay tetrahedra spanned by quartets of quasars, allowing for an estimate of the depth of underdense regions.  To maximise the BAO signal-to-noise ratio, we consider only voids with radii larger than 36$h^{-1}$Mpc. Our analysis shows a  negative BAO peak in the cross-correlation of QSOs and voids. The joint BAO measurement of the QSO auto-correlation and the corresponding cross-correlation with voids shows an improvement in 70$\%$ of the QSO mocks with an average improvement of $\sim5\%$. However, on the SDSS data, we find no improvement compatible with cosmic variance. For both mocks and data, adding voids does not introduce any bias. We find under the flat $\Lambda$CDM assumption, a distance joint measurement on data at the effective redshift $z_{\rm eff}=1.48$ of $D_V(z_{\rm eff})=26.297\pm0.547$. A forecast of a DESI-like survey with 1000 boxes with a similar effective volume recovers the same results as for light-cone mocks with an average of 4.8$\%$ improvement in 68$\%$ of the boxes.
\end{abstract}


%%%%%%%%%%%%%%%%%%%%%%%%%%%%%%%%%%%%%%%%%%%%%%%%%%

%%%%%%%%%%%%%%%%% BODY OF PAPER %%%%%%%%%%%%%%%%%%

\section{Introduction}
The accelerated expansion of the Universe is one of the greatest mysteries of current cosmology. It was observationally discovered  by \citet{Riess:1998aa} and \citet{Perlmutter:1999aa} a bit more than 20 years ago, but still its nature, referred to as dark energy, remains unknown. In the context of precision cosmology, an accurate determination of the expansion history of the Universe is required to constrain the nature of dark energy and thus to test the $\Lambda$CDM model.

To this goal, baryon acoustic oscillations (BAO) provide a characteristic length that enables measurement of the expansion rate \citep{Weinberg:2013aa}. BAO arises in the early Universe due to the counteracting plasma pressure and gravitation that produced sound waves. At photon decoupling, those waves stopped propagating, leaving an imprint detectable in the clustering of the galaxies and in the cosmic microwave background (CMB). The distance the waves travelled before they stopped, known as the sound horizon, can be used as a standard ruler \citep{Blake:2003aa}. 

The first BAO detections in the clustering of galaxies were made by \citet{Eisenstein:2005aa} with Sloan Digital Sky Survey (SDSS) data and \citet{Cole:2005aa} with Two Degree Field Galaxy Redshift Survey (2dFGRS). Since then, the era of spectroscopic surveys has risen with BAO as a key measurement. The largest survey to date is SDSS with Baryon Oscillation Spectroscopic Survey \citep[BOSS]{Dawson:2013aa} and at higher redshift with the extended Baryon Oscillation Spectroscopic Survey \citep[eBOSS]{Dawson:2016aa}. BAO was therefore measured at different redshifts in the clustering of various tracers such as luminous red galaxies \citep[LRGs;][]{Ross:2016aa, Bautista:2020aa, Gil-Marin:2020aa}, emission-line galaxies \citep[ELGs;][]{Raichoor:2020aa}, quasars \citep[QSOs;][]{Ata:2018aa} and Lyman-$\alpha$ forests \citep{Busca:2013aa,du-Mas-des-Bourboux:2020aa}.

\citet{Kitaura:2016aa} measured for the first time a BAO signal in the clustering of underdense regions, defined as voids. More recently, \citet{Zhao:2022aa} performed a multi-tracer with voids based on the analysis of ELG and LRG samples of BOSS and eBOSS. They showed that adding voids improved the BAO constraints of 5$\%$ to 15$\%$ for their samples (see also \citet{Zhao:2020ab}). Their studies relied on a Delaunay Triangulation \citep[DT;][]{Delaunay:1934aa} definition of voids (DT-voids), which detects a void as the largest empty sphere defined by four tracers \citep{Zhao:2016aa}. The voids are allowed to overlap, resulting in an increase of tracer number, which permits BAO detection, demarcating itself to other voids definitions used for redshift space clustering analysis \citep{Nadathur:2020ab, Aubert:2020aa}. 

At the precision level of current and future surveys like the Dark Energy Spectroscopic Instrument \citep[DESI]{DESI-Collaboration:2016aa,DESI-Collaboration:2016ab}, the 4-metre Multi-Object Spectroscopic Telescope \citep[4MOST]{deJong:2019aa} or Euclid \citep{Euclid:2011aa}, any reduction of measurement uncertainties will be crucial.

In this paper, we extend the work of \citet{Zhao:2022aa} by analysing the QSO sample of eBOSS using DT-voids. We provide a distance measurement from the joint BAO analysis of QSO auto-correlation and QSO-voids cross-correlation. The analysis pipeline and the errors are assessed using fast approximated mocks and N-body simulations. We also forecast error improvement from voids with a DESI-like survey for QSOs.

We summarise the QSO sample and the void catalogue used in Section~\ref{sec:data}. Fast mock catalogues and N-body simulations are introduced in Section~\ref{sec:mocks}. Method for void selection and correlation computation are described in Section~\ref{sec:method}. The BAO model and the template used for void fitting are outlined in Section~\ref{sec:model}. Error assessments are estimated in Section~\ref{sec:testsonmocks} and results in Section~\ref{sec:results} with our conclusions in Section~\ref{sec:conclusions}.

\section{Data}\label{sec:data}
We present in this section the eBOSS QSO sample used for the BAO analysis of this paper. We use the same QSO data catalogue as in the eBOSS DR16 analysis \citep{Hou:2020aa,Neveux:2020aa}, which was fully described in \citet{Ross:2020aa}.

The extended Baryon Oscillation Spectroscopic Survey \citep[eBOSS]{Dawson:2016aa} program was part of the fourth generation of the Sloan Digital Sky Survey \citep[SDSS-IV]{Blanton:2017aa} as an extension of the Baryon Oscillation Spectroscopic Survey \citep[BOSS]{Dawson:2013aa}. It aimed at observing the large-scale structure at higher redshifts. Started in 2014 until 2019, eBOSS used the double-armed spectrographs of BOSS \citep{Smee:2013aa} at the 2.5-meter aperture Sloan Telescope at Apache Point Observatory \citep{Gunn:2006aa}.

\begin{figure}
\centering
\includegraphics[width=0.95\columnwidth]{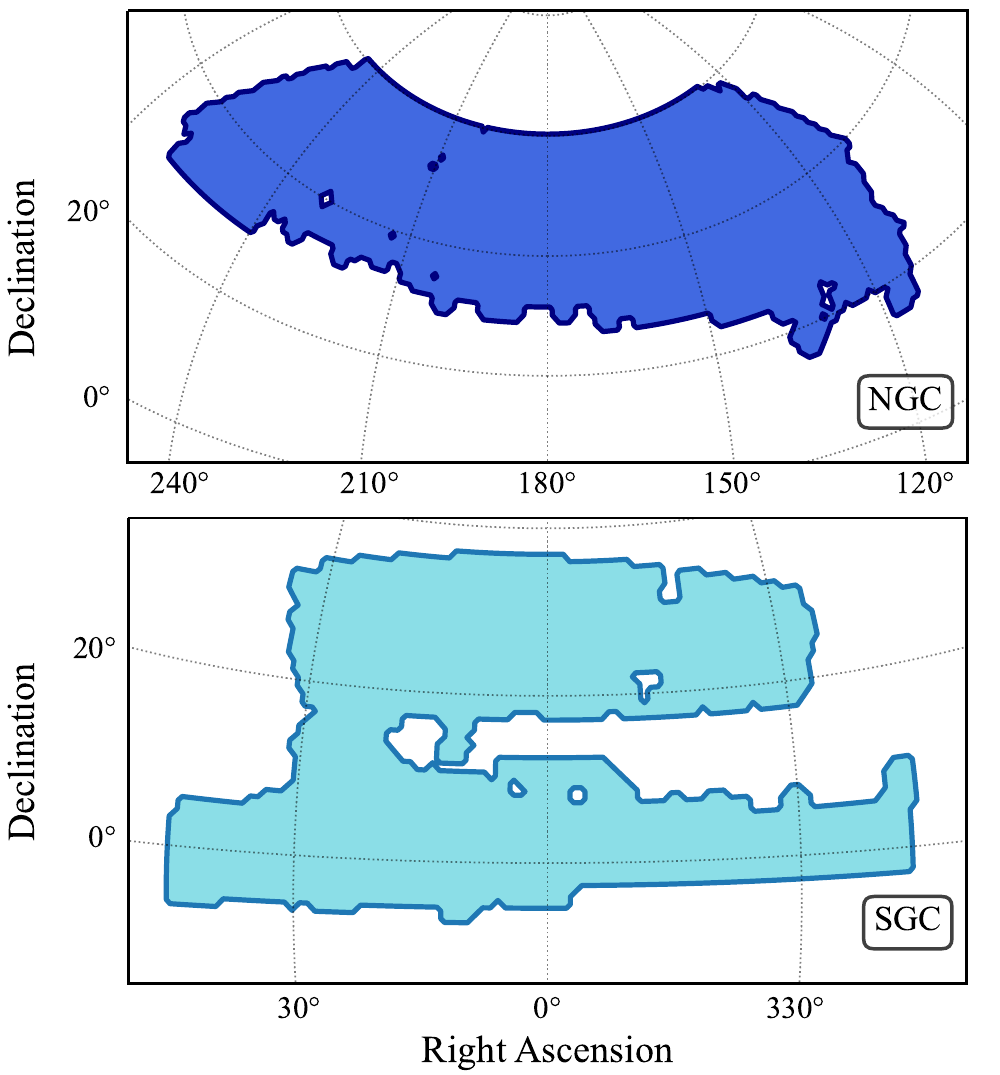}
\caption{\label{fig:footprint} 
Footprint of eBOSS DR16 QSO samples in the North (top) and South (bottom) Galactic Caps.}
\end{figure}

The eBOSS final release gathered reliable spectroscopic redshifts of over 340'000 QSOs in total, both in the South Galactic Cap (SGC) and North Galactic Cap (NGC), in a redshift range between 0.8 and 2.2. The QSOs were selected following the photometric target selection described in \citep{Myers:2015aa}. The footprints of both cap samples are presented in Figure~\ref{fig:footprint}. Different statistics as the weighted areas, the number of QSOs and the number densities are gathered in the Table~\ref{tab:QSO_stats}.

\iffalse
\begin{figure}
\centering
\includegraphics[width=1\columnwidth]{figures/Hist_nz_qso_voids.pdf}
\caption{\label{fig:nz} 
Redshift density of the eBOSS QSO and voids samples in the NGC (blue), SGC (cyan) and combined (red).}
\end{figure}
\fi

\begin{table}
	\centering
	\begin{tabular}{lccc}
		\hline 
		& NGC & SGC & Total\\
		\hline 
		Effective area [deg$^2$] & 2860 & 1839 & 4699 \\
		$\rm N_{\rm QSO}$ in 0.8 < $z$ < 2.2 & 218'209 & 125'499 & 343'708 \\
		$\rm n_{\rm QSO}$ [($h^{-1}\rm Mpc$)$^{-3}$] & $1.43\times10^{-5}$ & $1.60\times10^{-5}$ & $1.53\times10^{-5}$\\
        Effectif redshift & - & - & 1.48 \\
	    \hline
	\end{tabular} 
	\caption{Effective areas, effective redshift and number of reliable redshifts per Galactic cap and in the combined QSO sample in the redshit range 0.8 < $z$ < 2.2.}
	\label{tab:QSO_stats}
\end{table}

We apply weights to each individual QSO to account for observational and targeting systematics. We summarize here the different weights and refer to \citet{Ross:2020aa} for a complete description. The angular systematics due to the imaging quality is mitigated through the weight $w_{\rm sys}$. The weights $w_{\rm cp}$ and $w_{\rm noz}$ are respectively the close-pair  and redshift failure corrections. To minimize the clustering variance, we follow \citet{Feldman:1994aa} and apply the FKP weight $w_{\rm FKP}=(1+n(z)\cdot P_0)^{-1}$ where $n(z)$ is the weighted radial comoving number densities of QSO and $P_0=6000h^{-3}\rm Mpc^{3}$. The total weight applied to each QSO is then defined as their combination:

\begin{equation}
    w_{\rm tot} = w_{\rm sys}  \cdot w_{\rm cp} \cdot w_{\rm noz} \cdot w_{\rm FKP}.
\label{eq:weights}
\end{equation}

Following the eBOSS analyses, the QSO effective redshift $z_{\rm eff}$ is defined as the weighted mean of spectroscopic redshift over all galaxy pairs ($z_i,z_j$) in the separation range between 25 and 120 $h^{-1}$ Mpc:
\begin{equation}
    z_{\rm eff} = \frac{\sum_{i,j} w_{\rm{tot},i} w_{\rm{tot},j} (z_i + z_j)/2}{\sum_{i,j} w_{\rm{tot},i} w_{\rm{tot},j}}.
\label{eq:zeff}
\end{equation}

It gives for eBOSS QSO sample $z_{\rm eff}=1.48$, as presented in Table~\ref{tab:QSO_stats}.

A QSO random catalogue is built with about 50 times the QSO density. To account for the angular and radial distribution of the survey selection function, angular positions of random objects are uniformly drawn within the footprint, and their redshifts are randomly assigned from the data redshifts \citep{Ross:2020aa}. This radial selection introduces a radial integral constraint \citep[RIC]{de-Mattia:2019aa,Tamone:2020aa} which can affect the multipoles. It was shown in \citet{Hou:2020aa} and \citet{Neveux:2020aa} that this effect was relatively small for QSO.

\subsection{Void Catalogue}
The void data catalogue is constructed using the Delaunay Triangulation Void finder \citep[\textsc{DIVE}\footnote{\url{https://github.com/cheng-zhao/DIVE}}]{Zhao:2016aa}. It identifies the largest empty spheres formed by four distinct objects relying on the Delaunay triangulation \citep{Delaunay:1934aa} algorithm in comoving space. It provides the radii and centres of the empty spheres that we define as voids and take them as tracers. This definition allows the spheres to overlap, which permits a large number of objects and thus to detect a BAO peak allowing BAO measurements \citep{Kitaura:2016aa}. 

\begin{figure}
\centering
\includegraphics[width=0.9\columnwidth]{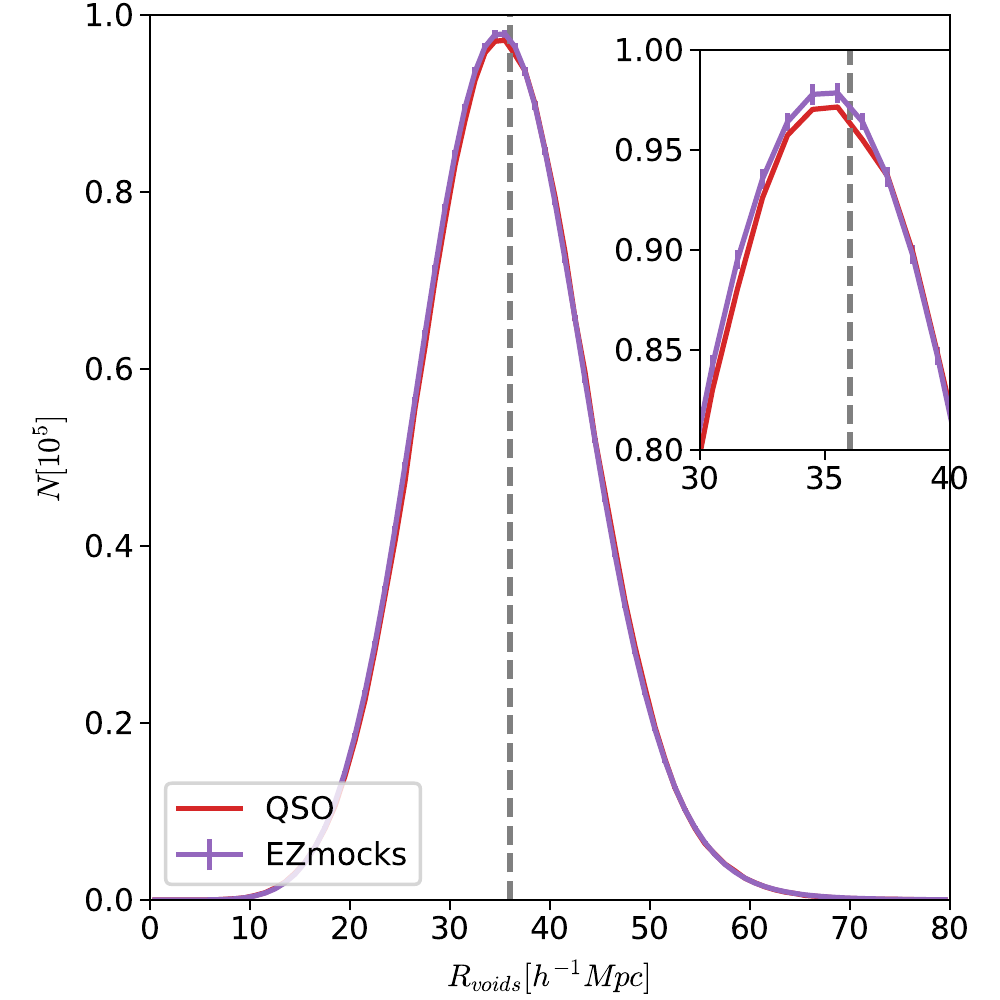}
\caption{\label{fig:nR} 
Radius of voids number density for the eBOSS QSO void sample and the EZmocks. Vertical line indicates the radius of 36 $h^{-1}$Mpc.}
\end{figure}

\begin{figure*}
\centering
\includegraphics[width=1.0\columnwidth]{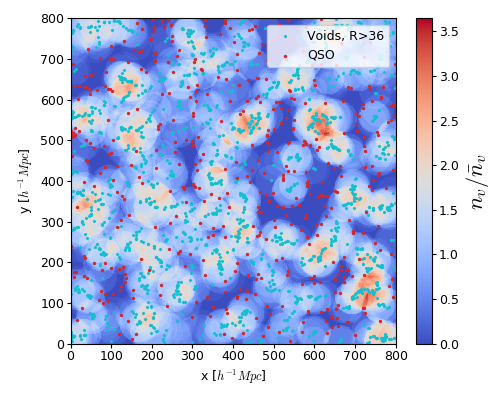}
\includegraphics[width=1.0\columnwidth]{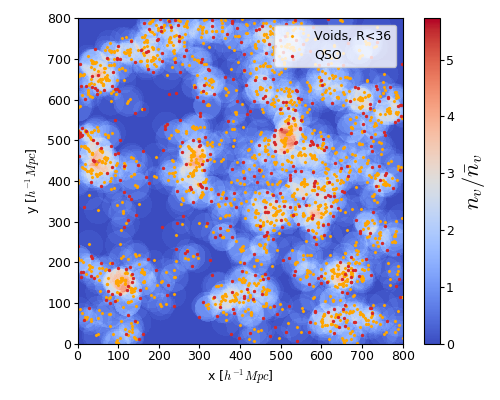}
\caption{\label{fig:QSOvR} Number density of spherical voids for a slice of NGC data sample of size 800$\times$800$\times$50 $h^{-3}$Mpc$^3$. QSOs are represented as red points. On the left: large voids, with radii larger than $36$ $h^{-1}$Mpc, center of voids are represented as red points. On the right: small voids, with radii smaller than $36$ $h^{-1}$Mpc, center of voids are represented as orange points. }
\end{figure*}

\textsc{DIVE} is run over the whole NGC and SGC data samples. The resulting voids are kept if their centre lies within the redshift range and footprints and outside the veto masks of the survey. The total number of voids is more than five times larger than the number of QSOs; see Table~\ref{tab:QSOvoids_stats}. The radius range of the voids displayed on Figure \ref{fig:nR}, spreads up to 80 $h^{-1}$Mpc with a mean radius around 35 $h^{-1}$Mpc. This is about twice the typical values obtained for LRGs and ELGs analysis with the same void definition \citep{Zhao:2020ab,Zhao:2022aa}. It can be easily explained due to the lower density of the QSO sample and the relationship between the number density and the size of the voids \citep{ForeroSanchez:2021aa}. Figure~\ref{fig:QSOvR} show QSOs and big (small) voids densities of a slice of NGC sample in comoving space. From them, one can see that the size of the voids is important: large voids track underdensities, while small voids lie in overdensity regions. These two populations of voids are respectively voids-in-voids and void-in-clouds \citep{Sheth:2004aa}. A careful choice of the radius of voids has to be made in order to avoid small voids contamination and therefore reduce the uncertainty of BAO measured from underdensities.

\begin{table}
	\centering
	\begin{tabular}{lccc}
		\hline 
		& NGC & SGC & Total\\
		\hline 
		$\rm N_{\rm voids}^{\rm tot}$ in 0.8 < $z$ < 2.2 & 1'304'614 & 718'966 & 2'023'580 \\
		$\rm N_{\rm voids}$ with 36 < $R$ < 80 & 589'549 & 373'362 & 962'911 \\
		$\rm n_{\rm voids}^{\rm tot}$ [($h^{-1}\rm Mpc$)$^{-3}$] & $8.18\times10^{-5}$ & $9.55\times10^{-5}$ & $9.01\times10^{-5}$\\
	    \hline
	\end{tabular} 
	\caption{Void number density and number of reliable redshifts per Galactic cap and in the combined QSO voids sample in the redshit range 0.8 < $z$ < 2.2.}
	\label{tab:QSOvoids_stats}
\end{table}

The random catalogues for voids are generated according to the procedure described in \citet{Liang:2016aa}. We stack 100 mock realizations and shuffle the angular positions and (redshift, radius) pairs within redshifts and radius bins of respectively  redshift 0.1 and 2 $h^{-1}Mpc$. We then randomly subsample down to 50 times the number of voids.

\section{Mocks}\label{sec:mocks}

We will introduce here different sets of mock catalogues used for this study. We work with approximate mocks to calibrate the data analysis pipeline and  estimate the covariance matrices. We use N-body simulations to validate the QSO-only BAO model.

\subsection{EZmocks}
EZmocks are fast approximated mocks relying on the Zel'dovich approximation \citep[ZA;][]{Zeldovich:1970aa}. The  displacement field of the ZA is generated from a Gaussian random field in a 5 $h^{-1}\rm Gpc$ box using a grid size of $1024^3$ with a given initial linear power spectrum. The dark matter density at the wanted redshift is then obtained by moving the dark matter particles directly to their final positions. Thereafter the simulation box is populated with QSOs using an effective galaxy bias model calibrated to the eBOSS DR16 QSO clustering measurements \citep{Chuang:2015aa,Zhao:2020aa}. It describes the relationship between the dark matter density field $\rho_{\rm m}$ and the tracer density field $\rho_{\rm t}$. This bias model \citep{Chuang:2015aa, Baumgarten:2018aa, Zhao:2020aa} requires a critical density $\rho_{\rm c}$ to form dark matter haloes \citep{Percival:2005aa}, an exponential cut-off $\rho_{\rm exp}$ \citep{Neyrinck:2014aa} and a density saturation $\rho_{\rm sat}$ for the stochastic generation of haloes.
\iffalse
\begin{equation}
\begin{split}
    & \rho_{\rm t} = B_s\theta(\rho_{\rm m}-\rho_{\rm c})\rho_{\rm sat}[1-\exp{-\frac{\rho_{\rm m}}{\rho_{\rm exp}}}], \\
    & {\rm where}\ B_s =
    \begin{cases}
      1 + X_{\lambda}, & \text{if $X_{\lambda}\ge0$}\\
      \exp{X_{\lambda}}, & \text{if $X_{\lambda}<0$}\\
    \end{cases}
    \ \ {\rm and}\ \theta(x) =
    \begin{cases}
      1, & \text{if $x\ge0$}\\
      0, & \text{if $x<0$}\\
    \end{cases},
\end{split}
\label{eq:biasEZ}
\end{equation}
with $X_{\lambda}$ a number generated from a gaussian distribution $\mathcal{N}(0,\lambda)$ centered at 0 with a standard deviation $\lambda$. Due to degeneracies with the other parameters, $\lambda$ and $\rho_{\rm sat}$ are both fixed to 10. 
\fi
The mocks are then populated following a probability distribution function (PDF) $P(n_{\rm t})=Ab^{n_{\rm t}}$, $n_{\rm t}$ being the number of tracers per grid cell, $b$ is a free parameter, and the parameter $A$ is constrained with the number density of QSOs in the box. Moreover the random motions are accounted for using a vector $X_{\nu}$ generated from a 3D gaussian distribution $\mathcal{N}(0,\nu)$, the peculiar velocity becomes: $u_{\rm t} = u_{\rm ZA} + X_{\nu}$, where $u_{\rm ZA}$ is the linear peculiar velocity in the ZA \citep{Bernardeau:2002aa}. In total we have 4 free parameters, namely $\rho_{\rm c}$, $\rho_{\rm exp}$, $b$ and $\nu$, that were calibrated to the data for the QSO eBOSS sample in \citet{Zhao:2020aa}.

The Flat-$\Lambda$CDM cosmology used for EZmocks is summarized in Table~\ref{tab:cosmo}.

For each different EZmocks set, we obtain a void catalogue by applying the same procedure than for the data.

\begin{table}
	\centering
	\begin{tabular}{lccc}
		\hline 
		& fiducial & EZmocks & OuterRim\\
		\hline 
		$h$ & 0.676 & 0.6777 & 0.71 \\
		$\Omega_m$ & 0.31 & 0.307115 & 0.26479 \\
		$\Omega_bh^2$ & 0.022 & 0.02214 & 0.02258 \\
		$\sigma_8$ & 0.8 & 0.8225 & 0.8 \\
		$n_s$ & 0.97 & 0.9611 & 0.963 \\
		$\sum m_{\nu}$ [eV] & 0.06 & 0 & 0 \\
	    \hline
	\end{tabular} 
	\caption{Different Flat-$\Lambda$CDM cosmologies used throughout the paper. Fiducial cosmology \citep{Planck-Collaboration:2016aa} is used for the template power spectrum and distance measurements for EZmocks and data. EZmock cosmology is the cosmology for EZmock creation. OuterRim cosmoly is the simulation cosmology and used for the fits to the N-body mocks.}
	\label{tab:cosmo}
\end{table}

\subsubsection{Cubic mocks}
We take directly 1000 EZmocks boxes that were used for the light-cone generation of the QSO eBOSS EZmocks \citep{Zhao:2020aa}. They are cubic boxes of 5 $h^{-1}\rm Gpc$ referred to as the {\it EZbox} all over this paper. They have at an effective redshift of $z=1.48$ and a number density of $n=2.4\cdot10^{-5}$ ($h^{-1}\rm Mpc$)$^{-3}$. We used them to determine the best radius cut of the QSO voids for this analysis. To this end we also produce a set of 200 EZbox without BAO at the effective redshift of the QSO sample using the same parameters than adopted in QSO eBOSS analysis\footnote{For the creation of the EZbox, we adopt parameters corresponding to $z=1.48$, the effective redshift of our sample, and with a number density of $n=2.4\cdot10^{-5}$ ($h^{-1}\rm Mpc$)$^{-3}$:  $(\rho_{\rm c},\ \rho_{\rm exp},\ b,\ \nu)=(0.4,\ 0.95,\ 0.003,\ 450)$.}.

The 1000 mocks with BAO included were given as input a linear matter power spectrum generated with the software \textsc{camb}\footnote{\url{https://camb.info/}} \citep{Lewis:2000aa}, while for the mocks without BAO, we use a linear power spectrum without wiggles generated following the model of \citet{Eisenstein:1998aa}. Both linear power spectra, with and without wiggles, are produced with the same set of cosmological parameters gathered as the EZmocks cosmology of Table~\ref{tab:cosmo}.

\iffalse
\begin{figure}
\centering
\includegraphics[width=1.0\columnwidth]{figures/densityDM_EBbox.png}
\caption{\label{fig:DMvoids} 
...}
\end{figure}
\fi

\subsubsection{Light-cones}

We use the same sets of light-cone EZmocks as the eBOSS DR16 analysis described in \citet{Zhao:2020aa} to evaluate the covariance matrices and to test the data analysis pipeline. They are constituted of 1000 realizations with systematics included for each cap, NGC and SGC.

To recreate the clustering evolution, each light-cone mock is built by combining seven snapshots at different redshifts sharing the same initial conditions. The survey footprint and veto masks are then applied to match the data geometry. 

Observational systematics effects from QSO data such as fibre collisions, redshift failure and photometric systematics are encoded into the EZmocks. Those effects are thereafter corrected by using some weights in the same way as for data (see Equation~\ref{eq:weights}). A random catalogue is produced for each EZmock with redshifts of the QSO catalogue assigned randomly.

\subsection{N-body simulations}

To assess the bias and tune the BAO model, we work with the N-body simulations built for the DR16 eBOSS analysis and described in \citet{Smith:2020aa}. They are produced from the \textsc{OuterRim} simulations \citep{Heitmann:2019aa} at a single redshift snapshot of $z=1.433$.

The \textsc{OuterRim} simulations are produced in a cubic box of 3 $h^{-1}\rm Gpc$ length with $10'240^3$ dark matter particles each with a mass of $m_p=1.82\cdot10^9$ $M_\odot h^{-1}$ using the WMAP7 cosmology \citep{Komatsu:2011aa} given in Table~\ref{tab:cosmo}. A Friends-of-Friends algorithm is used to detect dark matter haloes. The mocks are then populated with QSOs with 20 different halo occupation distribution (HOD) models and three different redshift smearing prescriptions described in \citet{Smith:2020aa}. Each different set is constituted of 100 realisations. In this paper, we will measure clustering, and BAO parameters on the 100 realisations of the 20 HOD mocks without smearing.

\section{Method}\label{sec:method}

This section presents details of the correlation function computation and the void selection.

\subsection{Two-point correlation functions}

To quantify the clustering of tracers in configuration space, we compute the two-point correlation function (2PCF) $\xi$ expressing the surplus of pairs separated by a vector distance $\boldsymbol{s}$ compared to a random uniform distribution. 

The observed redshifts are first converted into comoving distances using the same flat-$\Lambda$CDM fiducial cosmology as in eBOSS DR16 analysis, summarized in Table~\ref{tab:cosmo}. We then evaluate the pair counts of the different catalogues using the Fast Correlation Function Calculator (\textsc{FCFC}\footnote{\url{https://github.com/cheng-zhao/FCFC}}, Zhao in preparation). We compute for QSOs and voids the unbiased Landy–Szalay estimator of the isotropic 2PCF \citep[LS]{Landy:1993aa} for a pair separation of $s$: %and an angle $\theta$ between the line of sight and $\bf s$:
\begin{equation}
    \xi(s) = \frac{DD(s)-2DR(s,\mu)+RR(s)}{RR(s)},
\label{eq:LS}
\end{equation}
%$\mu=\cos\theta$, and
where $DD$, $DR$ and $RR$ are the normalized paircounts with $D$ denoting the tracer and $R$ the random catalogue.

For the cross-correlation (XCF) between QSOs, subscript q, and voids, subscript v, we use the following generalized estimator \citep{Szapudi:1997aa}:
\begin{equation}
    \xi_{\rm x}(s) = \frac{D_{\rm q}D_{\rm v}-R_{\rm q}D_{\rm v}-D_{\rm q}R_{\rm v}+R_{\rm q}R_{\rm v}}{R_{\rm q}r_{\rm v}}.
\label{eq:LScross}
\end{equation}
\iffalse
For the BAO analysis we compress the 2PCF into multipoles, in particular the monopole with $\ell=0$:
\begin{equation}
    \xi_{\ell}(s) = \frac{2\ell+1}{2}\int_{-1}^{1}\xi(s,\mu)P_{\ell}(\mu)d\mu,
\label{eq:multipoles}
\end{equation}
where $P_{\ell}$ are the Legendre polynomials with order $\ell$ \citep{Hamilton:1992aa}.
\fi
The two caps are combined into a single data sample for all the analysis by combining the paircounts \citep{Zhao:2022aa}:
\begin{equation}
\begin{split}
    & DD = \frac{n_{\rm SGC}^2DD_{\rm SGC}+n_{\rm NGC}^2DD_{\rm NGC}}{(n_{\rm SGC}+n_{\rm NGC})^2}, \\
    & DR = \frac{n_{\rm SGC}n_{\rm r,SGC}DR_{\rm SGC}+w_{\alpha}n_{\rm NGC}n_{\rm r,NGC}DR_{\rm NGC}}{(n_{\rm SGC}+n_{\rm NGC})(n_{\rm r,SGC}+w_{\alpha}n_{\rm r,NGC})}, \\
    & RR = \frac{n_{\rm r,SGC}^2RR_{\rm SGC}+w_{\alpha}^2n_{\rm r,NGC}^2RR_{\rm NGC}}{(n_{\rm r,SGC}+w_{\alpha}n_{\rm r,NGC})^2}.
\end{split}
\label{eq:combpaircounts}
\end{equation}
The weight $w_{\alpha}$ corrects for the different ratio data-random between the two sample, i.e. $w_{\alpha}=\frac{n_{\rm r,SGC}n_{\rm NGC}}{n_{\rm SGC}n_{\rm r,NGC}}$, and $n_{\rm i}$, $n_{\rm r, i}$ stand for the number of pairs in the data, random catalogues of the cap $\rm i$, respectively.

In the case of EZbox we use the natural estimator instead of the LS estimator which does not require a random catalogue: 
\begin{equation}
\xi(s)=\frac{DD(s)}{RR(s)_a} -1,
\end{equation}
where $RR_a=\frac{4}{3}\pi\left(s_{\rm max}^3-s_{\rm min}^3\right)\Delta\mu/L_{\rm box}$ is the analytical pair count for uniform randoms in a periodic box, with $L_{\rm box}$ the box length and $s_{\rm max}$, $s_{\rm min}$, $\Delta\mu$ are the separation bin boundaries.

\begin{figure}
\centering
\includegraphics[width=0.9\columnwidth]{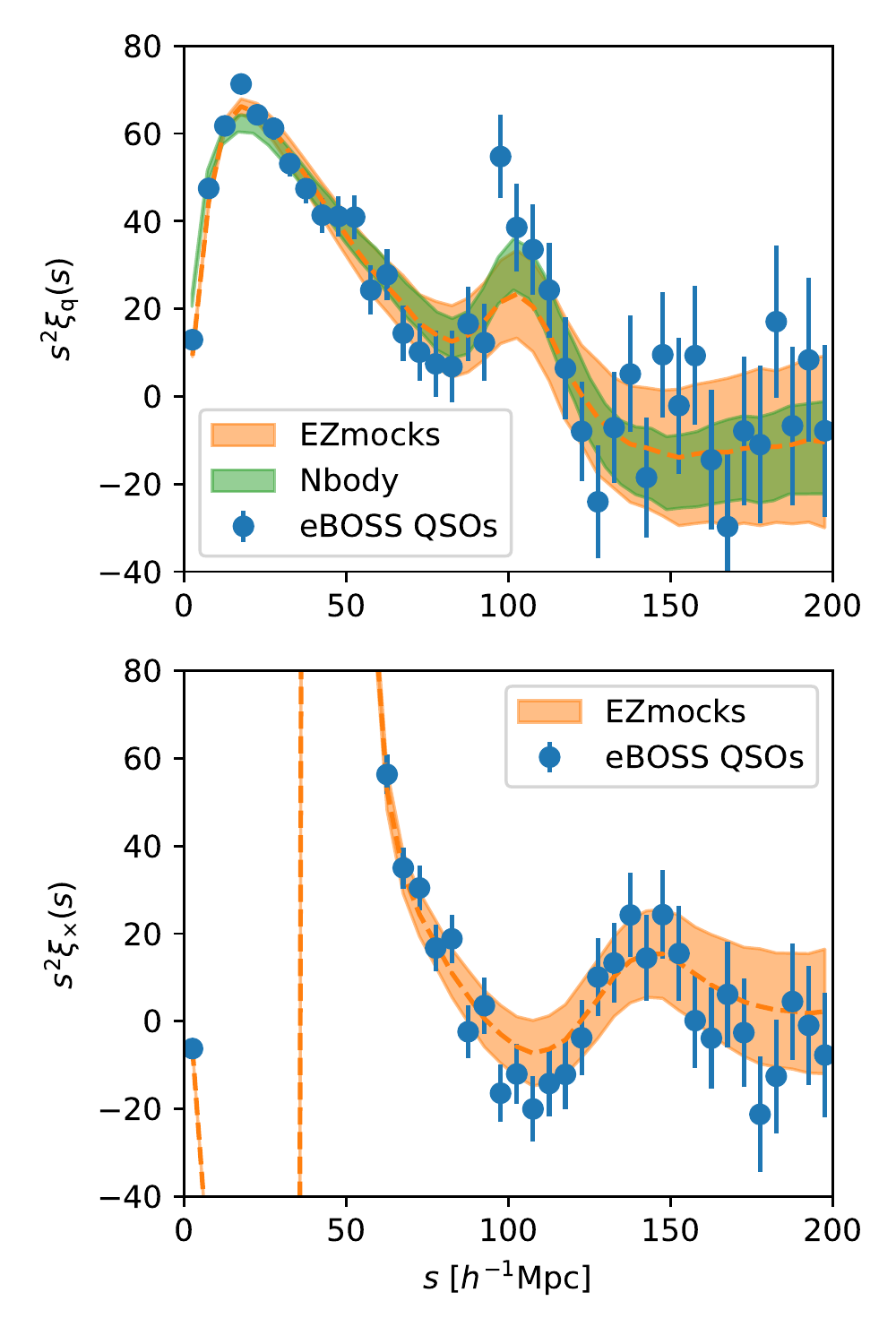}
\caption{\label{fig:CFeBOSSQSO} Top panel: auto-correlation of voids for the eBOSS QSOs sample with the standard deviation of EZmocks correlations as error bars. Mean of 1000 EZmocks  is in dashed line and their dispersion is in orange shaded area. Green shaded area indicates the mean of the 20x100 N-body simulations without smearing, rescaled to match EZmock cosmology (we rescaled with a factor $(r_{\rm d}^{\rm EZ}h^{\rm EZ})/(r_{\rm d}^{\rm OR}h^{\rm OR})=0.944$, 'OR' indicates the OuterRim cosmology, and 'EZ' EZmocks). Bottom panel: same for the cross-correlation of QSOs with QSOs voids larger than 36$h^{-1}$Mpc.}
\end{figure}

Figure~\ref{fig:CFeBOSSQSO} shows the auto-correlation of eBOSS QSO sample and its cross-correlation with QSOs large voids with a minimum void radius of 36$h^{-1}$Mpc.

\subsection{Covariances}

A covariance matrix $C$ is computed for each sample, i.e. QSOs auto-correlation and cross-correlation with voids, from the monopoles of 1000 EZmocks:
\begin{equation}
    C_{ij} = \frac{1}{N-1}\sum^N_{n=1}\left(\xi_0(s_i)-\frac{1}{N}\sum^N_{n=1}\xi_0(s_i)\right)\left(\xi_0(s_j)-\frac{1}{N}\sum^N_{n=1}\xi_0(s_j)\right),
\label{eq:cov}
\end{equation}
where $N$ is the total number of mocks and the subscripts $i,j$ run over the separation bins within the range considered. Those matrices are used to assess the errors of data and EZmocks. When the mean of the mocks is fitted, the covariance matrix is divided by $N$. For the multi-tracer covariance of 2PCF and XCF fitted jointly, the sum also runs over the cross-correlations of the two monopoles.

To obtain an unbiased estimator of the inverse covariance matrix $C^{-1}$, we multiply by the correction factor \citep{Hartlab:2007aa}, where $N_{d}$ is the number of separation bins used in the fit:
\begin{equation}
    C^{-1} = \left(1-\frac{N_b+1}{N-1}\right)C^{-1}.
\label{eq:cov}
\end{equation}

Analytical gaussian covariance matrices are computed following \citet{Grieb:2016aa} when fitting the QSO N-body mocks.

\subsection{Voids}
As mentioned previously, they are the two main populations of voids. The voids-in-clouds are tracers of overdensity regions, and voids-in-voids are tracers of underdense regions. These two types of voids can be set apart by their radius \citep{Zhao:2016aa}. \citet{ForeroSanchez:2021aa} showed that a constant radius cut gives a near-optimal signal-to-noise-ratio, SNR and that voids are less sensitive to observational systematics and therefore incompleteness. We chose to fix the maximum cut at $R_{\rm max}$=80$h^{-1}$Mpc to avoid contamination due to geometrical exclusion effects of very large voids, and we investigated the best minimum radius cut $R_{\rm min}$ that will be used in the analysis. 

\subsubsection{Correlation function}
\begin{figure*}
\centering
\includegraphics[width=1.0\columnwidth]{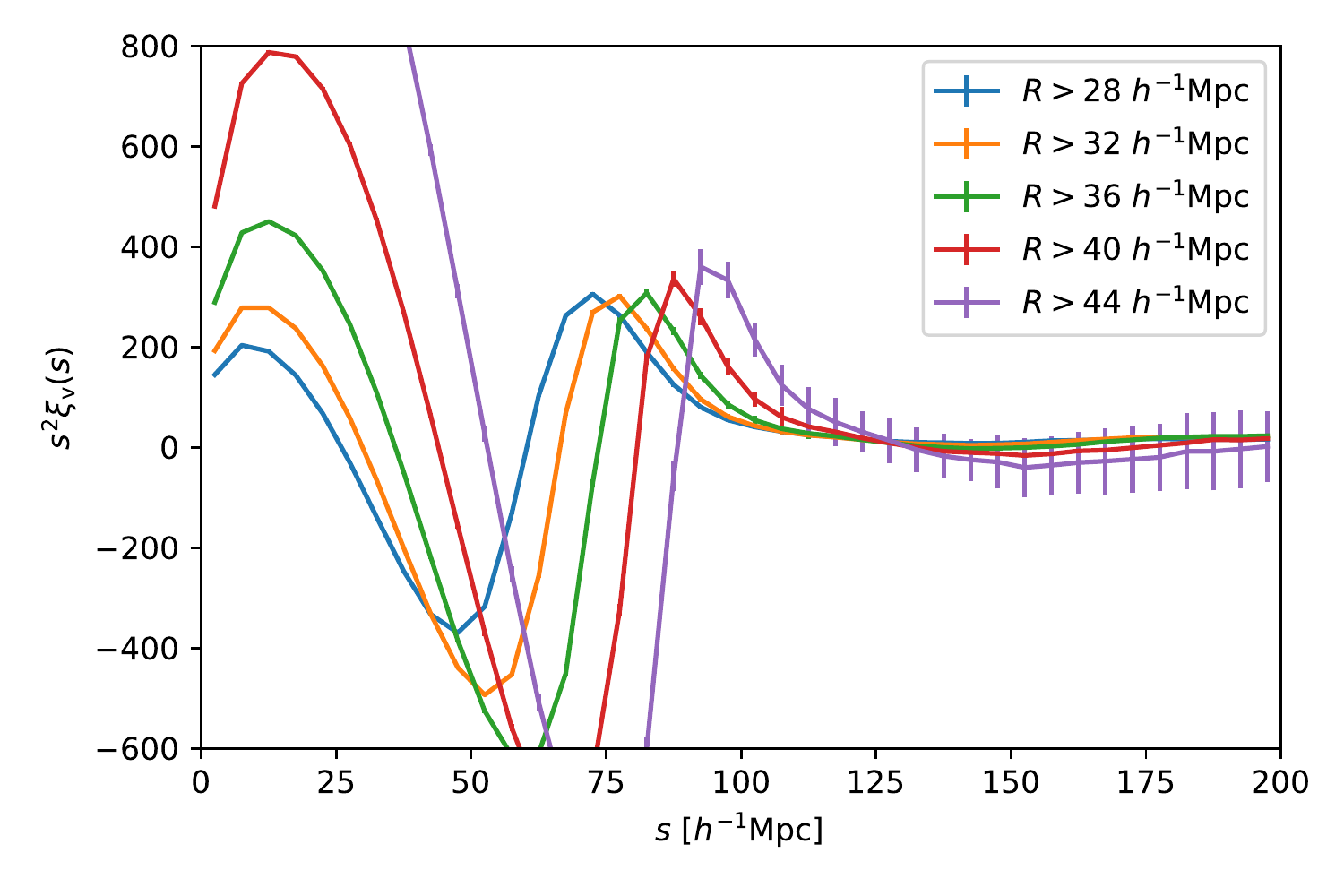}
\includegraphics[width=1.0\columnwidth]{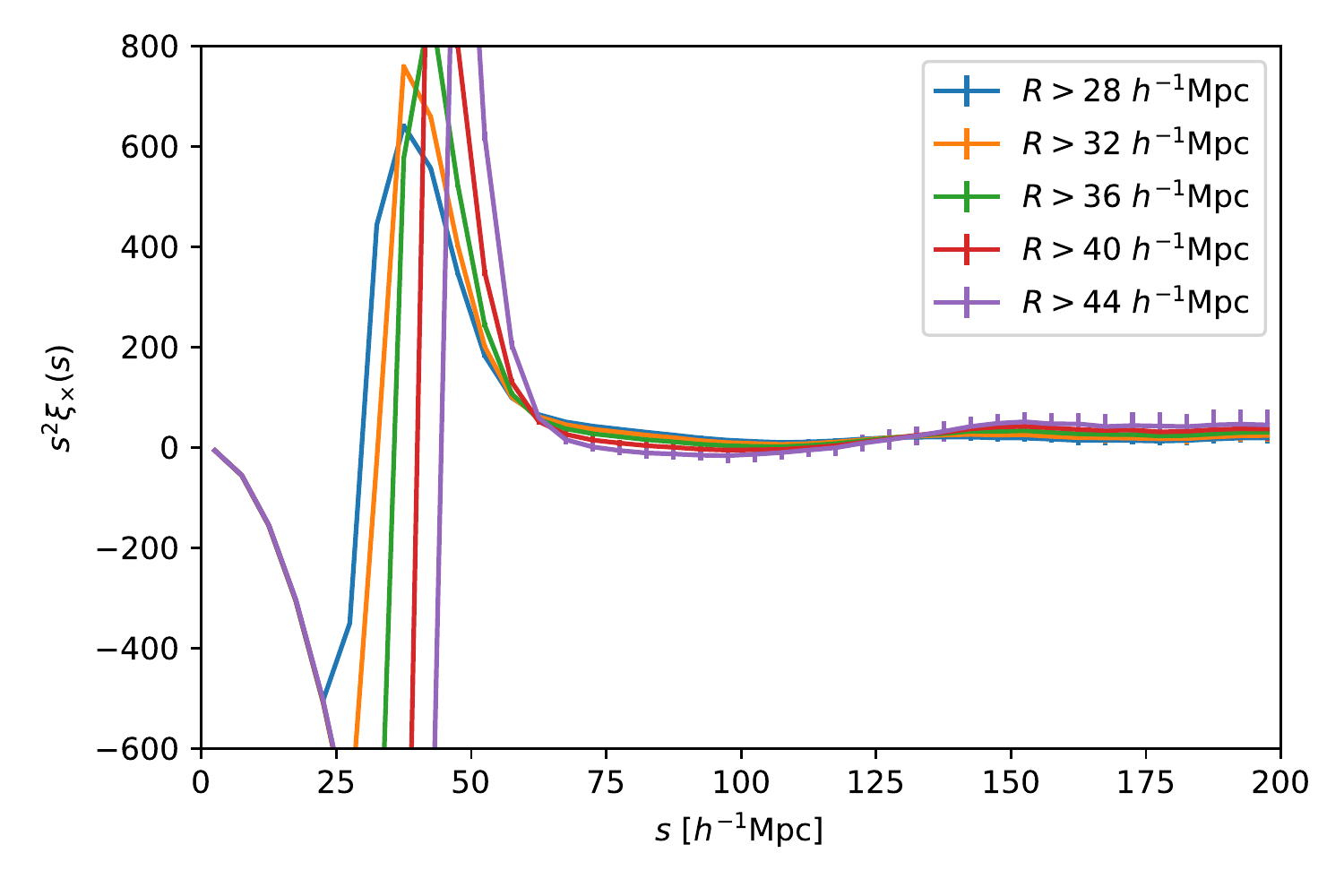}
\caption{\label{fig:corrEZvoidsRmin} Correlation functions for different radius cuts of the mean of 1000 EZmocks with standard deviation errors. Radius range is from $R_{\rm min}=R$ to $R_{\rm max}=80$ $h^{-1}$Mpc. On the left: auto-correlation of QSOs voids. On the right: cross-correlation of QSOs and voids. }
\end{figure*}

Correlation functions for different radius cuts are shown on Figure~\ref{fig:corrEZvoidsRmin} for QSO eBOSS EZmocks. The auto-correlation of voids (left panel of Figure~\ref{fig:corrEZvoidsRmin}) presents a very strong exclusion pattern, similar to what is observed for haloes due to their finite size \citep{Sheth:1999ab,Baldauf:2013aa}. Indeed even though the DT voids are not distinct from each other and can overlap, there is still an exclusion effect due to finite void size geometry \citep{Chan:2014aa,Zhao:2016aa}. As the minimum radius cut required to have large enough voids is about twice the value for LRG, see \citet{Zhao:2020ab} and \citet{Zhao:2022aa}, the exclusion effect due to the spherical definition of the voids is therefore also shifted to the right. It implies that the exclusion pattern interferes with the BAO scale. Around 100 $h^{-1}$Mpc, the correlation is noisy, and the BAO excess density is not detectable due to the strong signal of the void exclusion. This is why we chose in this paper to leave aside the auto-correlation of voids in the analysis and concentrate on their cross-correlation with QSOs.

On the right panel of Figure~\ref{fig:corrEZvoidsRmin} is the cross-correlation of QSOs with voids cut at different minimum radius $R_{\rm min}$ for EZmocks. The exclusion effect is still present, but it mainly affects scales up to twice the minimum radius $R_{\rm min}$. Therefore it has fewer effects on the BAO scale even though this is not obvious to understand its real effect. We refer to the next section for analysis of non-wiggles boxes to quantify this effect.

\begin{figure*}
\centering
\includegraphics[width=1.0\columnwidth]{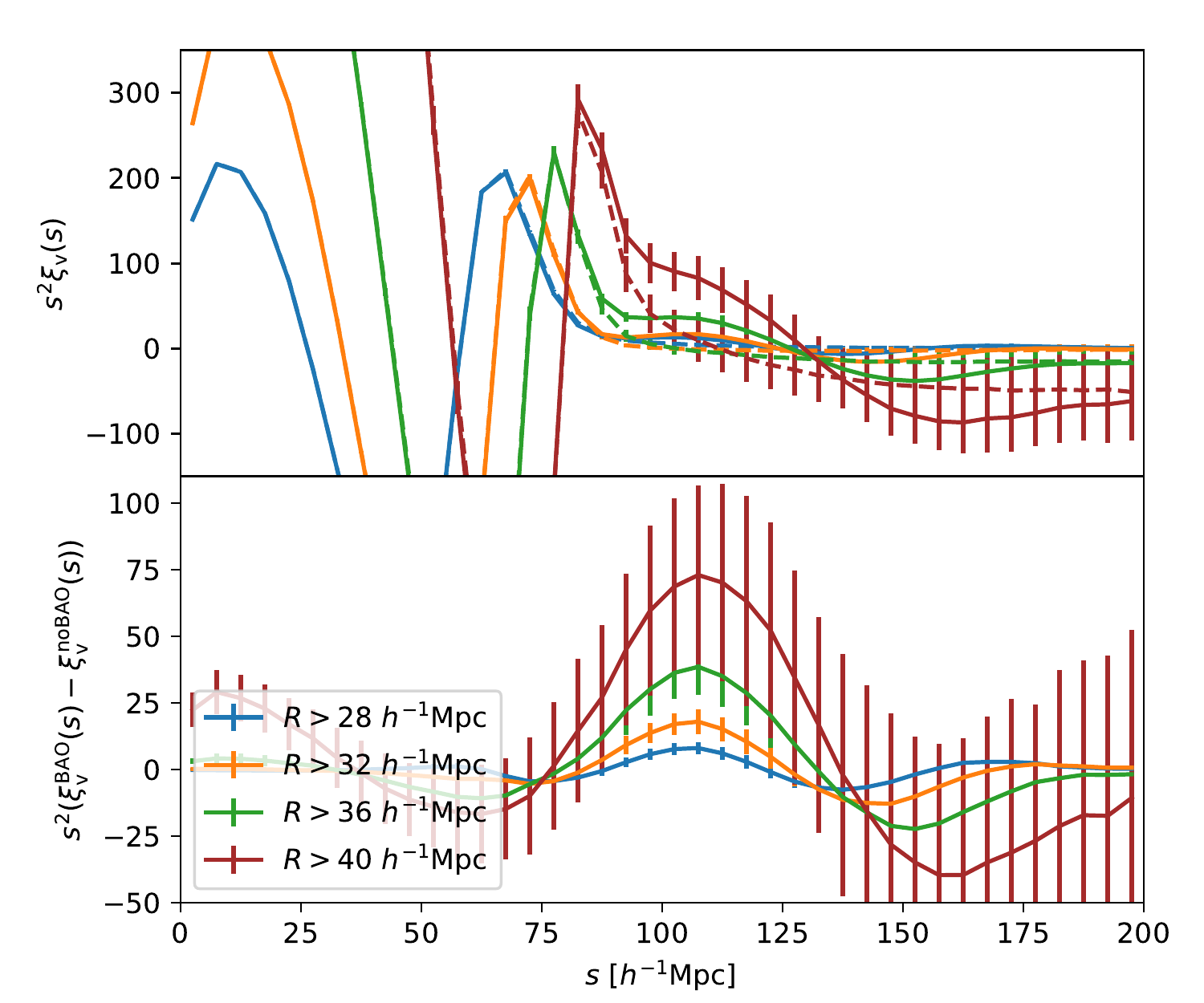}
\includegraphics[width=1.0\columnwidth]{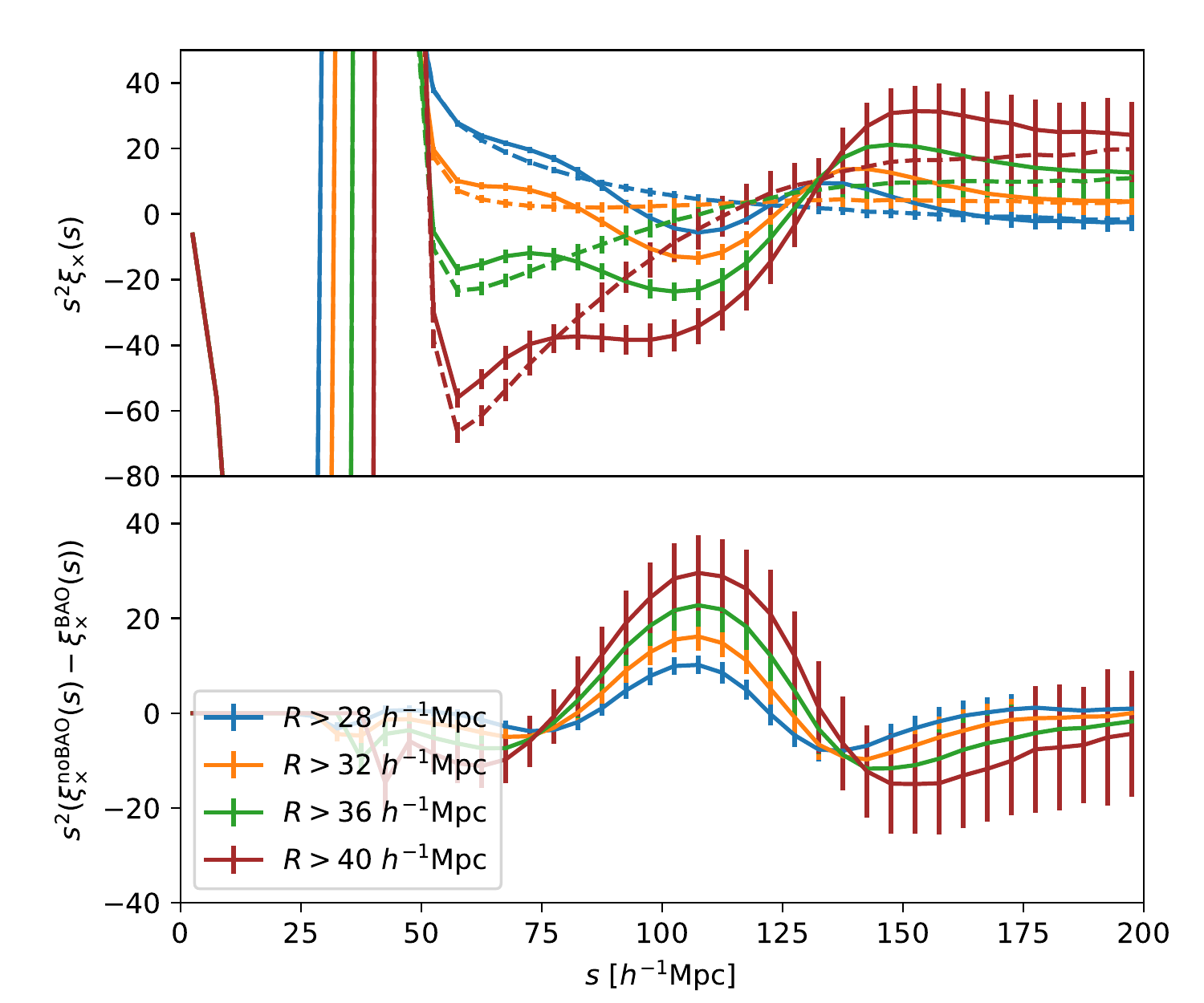}
\caption{\label{fig:corrEZbox} On the left, top panel: auto-correlation of QSO voids of the mean of 200 EZbox for a radius range from $R_{\rm min}=R$ to $R_{\rm max}=80$ $h^{-1}$Mpc, with standard deviation errors. Solid lines are for EZbox with BAO, dashed lines are for EZbox without BAO. Bottom left panel: mean difference of the auto-correlation of EZbox without BAO and EZbox with BAO, for different radius cut. On the right figures, same but for the cross-correlation of QSOs and voids.}
\end{figure*}

\subsubsection{Selection of optimal radius}\label{EZboxR}
To understand the exclusion effect on the cross-correlation of voids and QSOs at the BAO scale and to find a quantifiable way to select the optimal radius, we rely on the EZbox produced with and without BAO.

The top left (right) panel of Figure~\ref{fig:corrEZbox} displays the void auto (cross)-correlation of EZbox with and without BAO. In the cross-correlation, a net negative peak around 100$h^{-1}$Mpc can be seen from the BAO mocks compared to the ones without BAO wiggles. The bottom panels of Figure~\ref{fig:corrEZbox} show the difference between the two kinds of mocks, i.e. $\xi^{\rm no\ BAO} - \xi^{\rm BAO}$, another way to see the BAO excess that manifests itself as a clear bump. While we understand from the plots that a BAO peak is detectable from the void auto-correlation as well, we still chose not to include it in the analysis to avoid contamination from the exclusion effect in the model. Indeed if the exclusion effect is not perfectly modelled, the BAO fitting results might be biased.

To select the optimal radius threshold, we determine an SNR different to what was used in previous studies with DT voids \citep{Liang:2016aa}. We rely on the EZbox for the SNR computation and compute the area $A$ between the two EZbox curves over a selected separation range $S$ around the BAO peak:
\begin{equation}
    A = \sum_{s_i\in S} \xi_0^{\rm no\ BAO}(s_i)-\xi_0^{\rm BAO}(s_i).
\label{eq:A}
\end{equation}
For a radius cut $R_{\rm min}$, the signal $S_A$ is then defined as the mean of $A$ and the noise $N_A$ as the standard deviation of $A$ over the 200 EZbox. The SNR is $S_A/N_A$. 

The BAO signal and noise both increase with the minimum radius, as the underdense regions are better selected, but the total number of retained voids decreases. We observe a slight shift of the BAO peak to the larger scale that we understand as remaining exclusion effects that spread on the BAO scale.

\begin{figure}
\centering
\includegraphics[width=0.7\columnwidth]{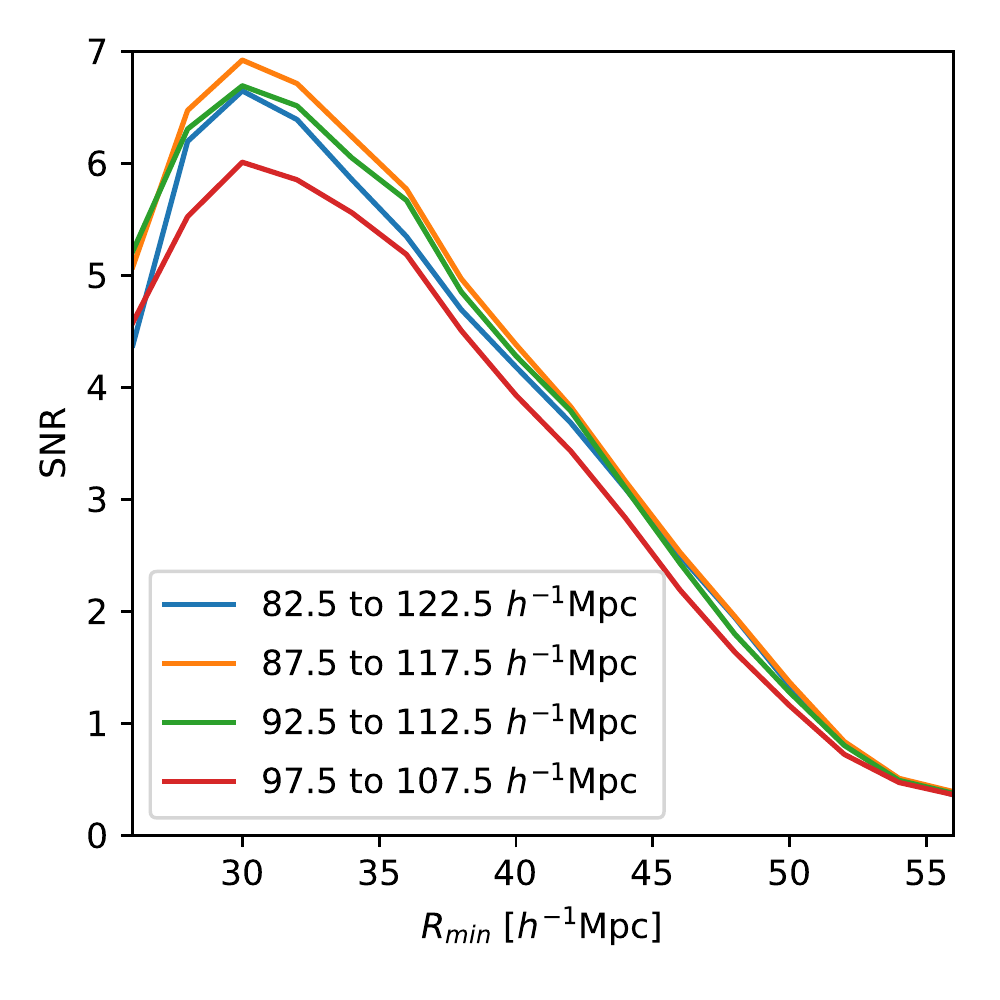}
\caption{\label{fig:SNR} Signal-to-noise ratio $SNR$ as defined with Equation~\ref{eq:A} as a function of the minimum radius $R_{\rm min}$. Different curves are for different separation range $S$ for which the minimum and maximum separation is indicated in the legend in $h^{-1}$Mpc.}
\end{figure}

We compute the SNR for different radius cuts over different separation ranges $S$, as shown in Figure~\ref{fig:SNR}. The optimal ratio featuring the higher SNR for all $S$ definitions is 31$h^{-1}$Mpc. It corresponds to the quantile of the void radius distribution of about 0.55. Reporting this quantile from EZbox to data and EZmocks gives:
\begin{equation}
    R_{\rm min}^{\rm optimal} =36 h^{-1}\rm Mpc.
\label{eq:Rmin}
\end{equation}
We chose, therefore, this value as the optimal minimum radius cut for our analysis of EZmocks and data. The number of voids with this radius cut is presented in Table~\ref{tab:QSOvoids_stats}. There are a bit less than three times more voids than QSOs.

\section{Model}\label{sec:model}

Here we present the models for the two-point statistics to extract the BAO signature for the voids and QSOs.

\subsection{Isotropic BAO}
The BAO peak in the clustering of the tracers, positive for big voids and QSOs auto-correlations and negative for their cross-correlation, is shifted if a wrong cosmology is assumed when transforming redshifts to distances. This effect is known as the Alcock-Paczynski (AP) effect \citep{Alcock:1979aa}. We account for the AP effect with the isotropic AP dilation parameter $\alpha$:
\begin{equation}
\alpha=\frac{D_Vr_{\rm d,fid}}{D_{V,\rm fid}r_{\rm d}}.
\label{eq:alpha}
\end{equation}
Subscript 'fid' stands for fiducial values used in the analysis. Parameter $r_{\rm d}$ is the comoving sound horizon at the baryon drag epoch when the baryon optical depth is one \citep{Hu:1996aa}, and $D_V$ is a volume-averaged distance defined as:
\begin{equation}
    D_V = \left(D_M(z)^2\frac{cz}{H(z)}\right)^{\frac{1}{3}},
\label{eq:DV}
\end{equation}
with $D_M$ the comoving angular diameter distance, $H(z)$ the Hubble parameter at redshift $z$, and $c$ the speed of light \citep{Eisenstein:2005aa}.

The theoretical BAO model $\xi_{\rm m}$ for the correlation that we use is:
\begin{equation}
\xi_{\rm m}(s) = B\xi_{\rm temp}(\alpha s) + A_0 + A_1/s + A_2/s^2,
\label{eq:xim}
\end{equation}
where $B$ is the tracer bias, controlling the amplitude,  and the $A_i$ with $i=0,1,2$ are broadband parameters treated as nuisance parameters. The model relies on a 2PCF template $\xi_{\rm temp}$ which is the Fourier transform of the power spectrum $P_{\rm temp}$:
\begin{equation}
\xi_{\rm temp}(s) = \frac{1}{2\pi^2}\int P_{\rm temp}(k)j_0(ks)e^{-k^2a^2}k^2dk.
\label{eq:xitemp}
\end{equation}
The function $j_0$ is the Bessel function at order 0 of the first kind. Here, the $a$ parameter is damping the high $k$ oscillations and is fixed at 2 $h^{-1}$Mpc following \citet{Variu:2022aa}. Indeed they demonstrate that BAO measurements are unbiased and more robust against template noise with $a$=2 $h^{-1}$Mpc compared to smaller values. The template power spectrum $P_{\rm temp}$ is \citep{Xu:2012aa}:
\begin{equation}
P_{\rm temp}(k) = \left( P_{\rm lin}(k) - P_{\rm lin, nw}(k) \right) e^{-k^2\Sigma_{\rm nl}^2/2} + P_{\rm lin, nw}(k),
\label{eq:Ptemp}
\end{equation}
where $\Sigma_{\rm nl}$ is the BAO damping parameter of the tracer, $P_{\rm lin}$ and $P_{\rm lin, nw}$ are the linear matter power spectrum and its analogue without BAO wiggles, respectively, produced in the same way as for EZbox using the fiducial cosmology of Table~\ref{tab:cosmo}.

\subsection{De-wiggled BAO model}
The de-wiggled template BAO model is not accurate for voids correlation functions \citep{Zhao:2020ab} because of oscillatory patterns inserted in power spectra due to void exclusion \citep{Chan:2014aa}. Equation \eqref{eq:Ptemp} is then modified to try to correct for this effect as:
\begin{equation}
P_{\rm t}(k) = P_{\rm temp}(k) \frac{P_{\rm tracer, nw}(k)}{P_{\rm lin, nw}(k)}.
\label{eq:template}
\end{equation}
The term $P_{\rm tracer, nw}(k)$ is the non-wiggle power spectrum of the tracer encoding broad-band and geometric effects. Those effects for DT voids are difficult to model. In a previous analysis study with voids, a parabolic parametrisation was introduced with an additional free parameter \citep{Zhao:2020ab, Zhao:2022aa} to model the non-wiggle ratio. However, this method does not work well for QSOs voids correlation as the exclusion effect is much stronger. This is why in this study, we rely on the second method, which is template-based \citep{Zhao:2022aa, Variu:2022aa}.

Developed by \citet{Variu:2022aa} with the Cosmological GAussian Mock gEnerator (\textsc{CosmoGAME}\footnote{\url{https://github.com/cheng-zhao/CosmoGAME}}), the de-wiggles tracer template is constructed with mocks without BAO wiggles. Those are Lagrangian mocks built on a Gaussian random field generated from $P_{\rm lin, nw}(k)$, with a simple galaxy bias selection tuned to match eBOSS QSO EZmocks. Survey geometry and radial selection are then applied to the mock catalogues. 

\begin{figure}
\centering
\includegraphics[width=0.49\columnwidth]{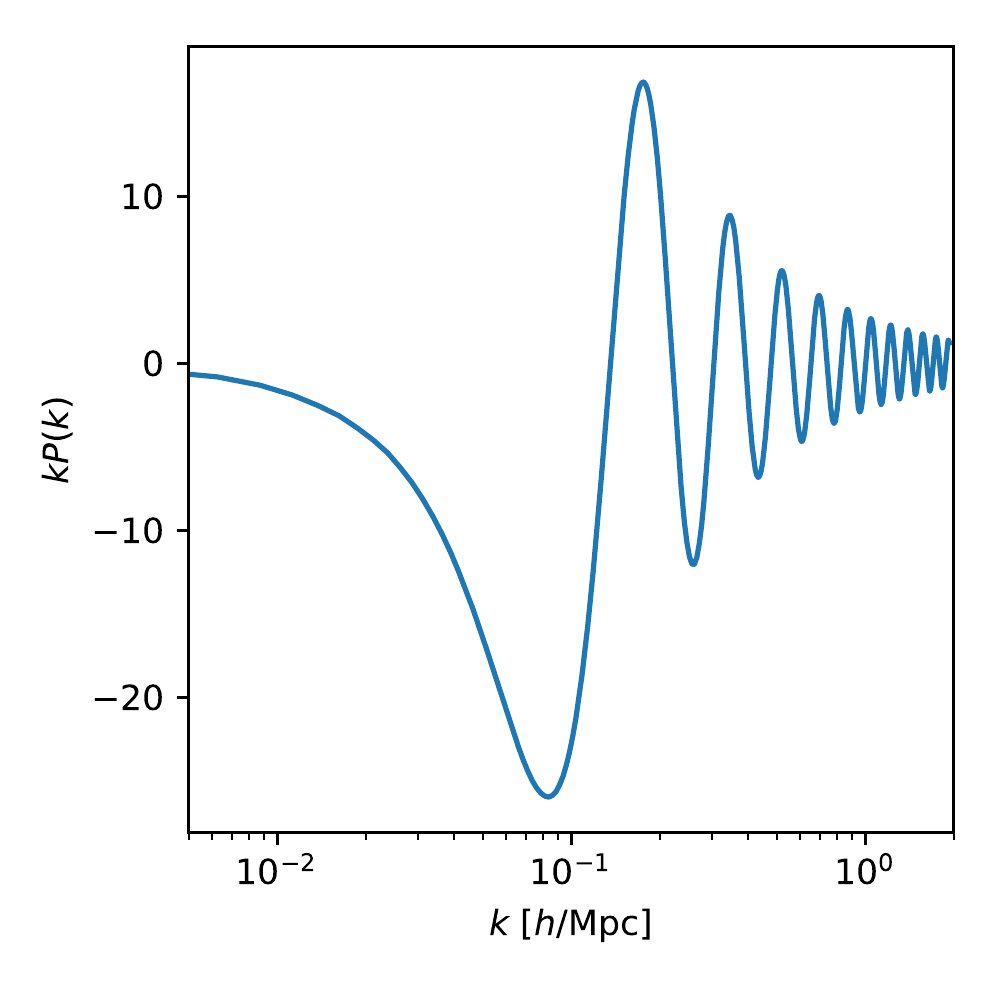}
\includegraphics[width=0.49\columnwidth]{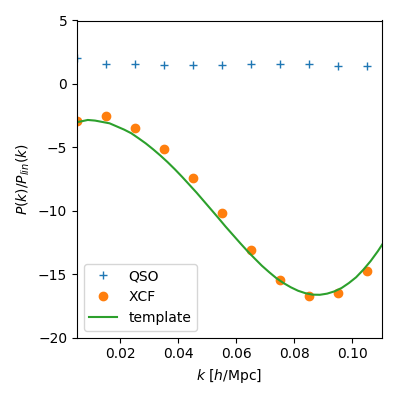}
\caption{\label{fig:template} Left pannel: De-wiggle template for the cross-correlation of QSOs and voids for the eBOSS QSO sample, generated with \textsc{CosmoGAME}. Right pannel: Power spectrum divided by the linear power spectrum $P_{\rm lin}(k)$. Blue crosses is for the mean QSOs auto-power spectrum of 100 EZmocks. Orange points is the QSOs and voids cross-power spectrum of 100 EZmocks. Solid green line is the cross-correlation template rescaled roughly to match the cross-correlation.}
\end{figure}

The template for the cross-correlation of QSOs and voids is obtained by averaging and stacking 2000, 1000, 100 mocks generated with CosmoGAME over a k-range $k$ up to 0.3, 1, 2 $h$Mpc$^{-1}$, respectively. Their power spectra are computed with \textsc{Powspec}\footnote{\url{https://github.com/cheng- zhao/powspec}}. The resulting concatenated template is shown in Figure~\ref{fig:template}, and its comparison with the power spectrum from 100 EZmocks is on the right panel of Figure~\ref{fig:template}.

\subsection{Parameter estimation}
To obtain BAO constrain we use the algorithm \textsc{Multinest}\footnote{\url{https://github.com/farhanferoz/MultiNest}} \citep{Feroz:2009aa} and its python version \textsc{pyMultinest}\footnote{\url{https://github.com/JohannesBuchner/PyMultiNest}} \citep{Buchner:2014aa},  an efficient Monte-Carlo method that computes Bayesian evidence and produce posteriors. We use the following likelihood assuming the gaussianity of the distribution for a given set of parameters $p$:
\begin{equation}
L\propto \exp{\left(-\chi^2(p)/2\right)},
\end{equation}
where the chi-scared function $\chi^2(p)$ is computed from the data $\xi_{d}$ and the model prediction depending on the parameter set $p$, $\xi_{\rm temp}(p)$:
\begin{equation}
\chi^2(p)=\left(\xi_{d}-\xi_{\rm temp}(p)\right)^{T}C^{-1}\left(\xi_{d}-\xi_{\rm temp}(p)\right).
\end{equation}

The resulting parameter covariances are rescaled to correct for the covariance matrix uncertainty propagation by \citet{Percival:2014aa}:
\begin{equation}
m_1 = \frac{1+(N_{\rm d}-N_{\rm par})\cdot B}{1+A+(1+N_{\rm par})\cdot B},
\label{eq:m1}
\end{equation}
where $N_{\rm d}$ the total number data bins used in the fit with $N_{\rm par}$ free parameters, and $A$ and $B$ are ($N_{\rm m}$ is the number of mocks used to estimate the covariance):
\begin{equation}
A = \frac{2}{(N_{\rm m}-N_{\rm d}-1)\dot(N_{\rm m}-N_{\rm d}-4)},
\end{equation}
\begin{equation}
B = \frac{N_{\rm m}-N_{\rm d}-2}{(N_{\rm m}-N_{\rm d}-1)\cdot(N_{\rm m}-N_{\rm d}-4)}.
\end{equation}
Distribution variance of multiples best-fits values from mocks used for the covariance has to be rescaled by:
\begin{equation}
m_2 = \left(1-\frac{N_{\rm d}+1}{N_{\rm m}-1}\right)m_1.
\label{eq:m2}
\end{equation}

The parameter set for the multi-tracer analysis of the auto-correlation of QSOs and their cross-correlation with voids is: $p=\left(\alpha,B_{\rm q},B_{\rm \times},\Sigma_{\rm nl, q},\Sigma_{\rm nl,\times}\right)$. In the single tracer analysis, only one $B$ and $\Sigma_{\rm nl}$ are used. Fits are performed with the BAO Fitter for muLtI-Tracers (\textsc{BAOflit}\footnote{\url{https://github.com/cheng-zhao/BAOflit}} code from \citet{Zhao:2022aa}. When let free, we chose very wide priors for each parameter, it corresponds to the first row of Table~\ref{tab:priors}. Broad-band parameters $A_i$ of the polynomial term in Equation~\ref{eq:xim} are determined by linear regression with the least squares method.

\begin{table}
    \centering
    \resizebox{\columnwidth}{!}{ 
    \begin{tabular}{lccccc}
    \hline
         & $\alpha$ & $B_{\rm q}$ & $\Sigma_{\rm nl, q}$ [Mpc/$h$] & $B_{\rm \times}$ & $\Sigma_{\rm nl, \times}$ [Mpc/$h$]\\
    \hline
        Flat & 0.8-1.2 & 0-100 & 0-100 & 0-100 & 0-100 \\
        $\xi_{\rm q}$ & 0.8-1.2 & 1.27-1.40 & 5.2 (6.7) & - & -\\
        $\xi_{\rm mt}$ & 0.8-1.2 & 1.27-1.40 & 5.2 (6.7) & 8.22-9.68 & 12.9 \\
    \hline
    \end{tabular}
    }
    \caption{ Prior ranges of the BAO bayesian analysis for the three parameters $\alpha$, $B$, $\Sigma_{\rm nl}$. Top row is for free parameters. Other rows are our fiducial choices when fitting the 2PCF or in the multi-tracer case. Value in parenthesis for $\Sigma_{\rm nl,q}$ is the value used when fitting EZmocks.}
    \label{tab:priors}
\end{table}

\section{Tests on mocks}\label{sec:testsonmocks}
We use eBOSS EZmocks to test the pipeline, calibrate the different settings for the analysis of data and assess systematics. N-body mocks are also used when dealing with QSOs only. We fit the auto-correlations of QSOs $\xi_{\rm q}$ (with Equation~\ref{eq:Ptemp}) and the cross-correlations with voids $\xi_{\times}$ (with Equation~\ref{eq:template}) first separately, and then we perform a multi-tracer fit where both correlations are fitted simultaneously, noted $\xi_{\rm mt}\equiv\{\xi_{\rm q}, \xi_\times \}$. Voids used are selected by the criterion in Equation~\ref{eq:Rmin}.

\subsection{Fitting ranges}
To choose our fiducial separation fit ranges, we fit the mean of the 1000 EZmocks for the QSO auto-correlation and cross-correlation, varying the fitting range. We aim to extract the maximum information and reduce the errors. Covariance matrices are divided by the number of mocks $N_m$ used to construct it, i.e. rescaled by 0.001. All the parameters are let free, i.e. with broad enough priors of Table~\ref{tab:priors}.

\begin{figure}
\centering
\includegraphics[width=1.0\columnwidth]{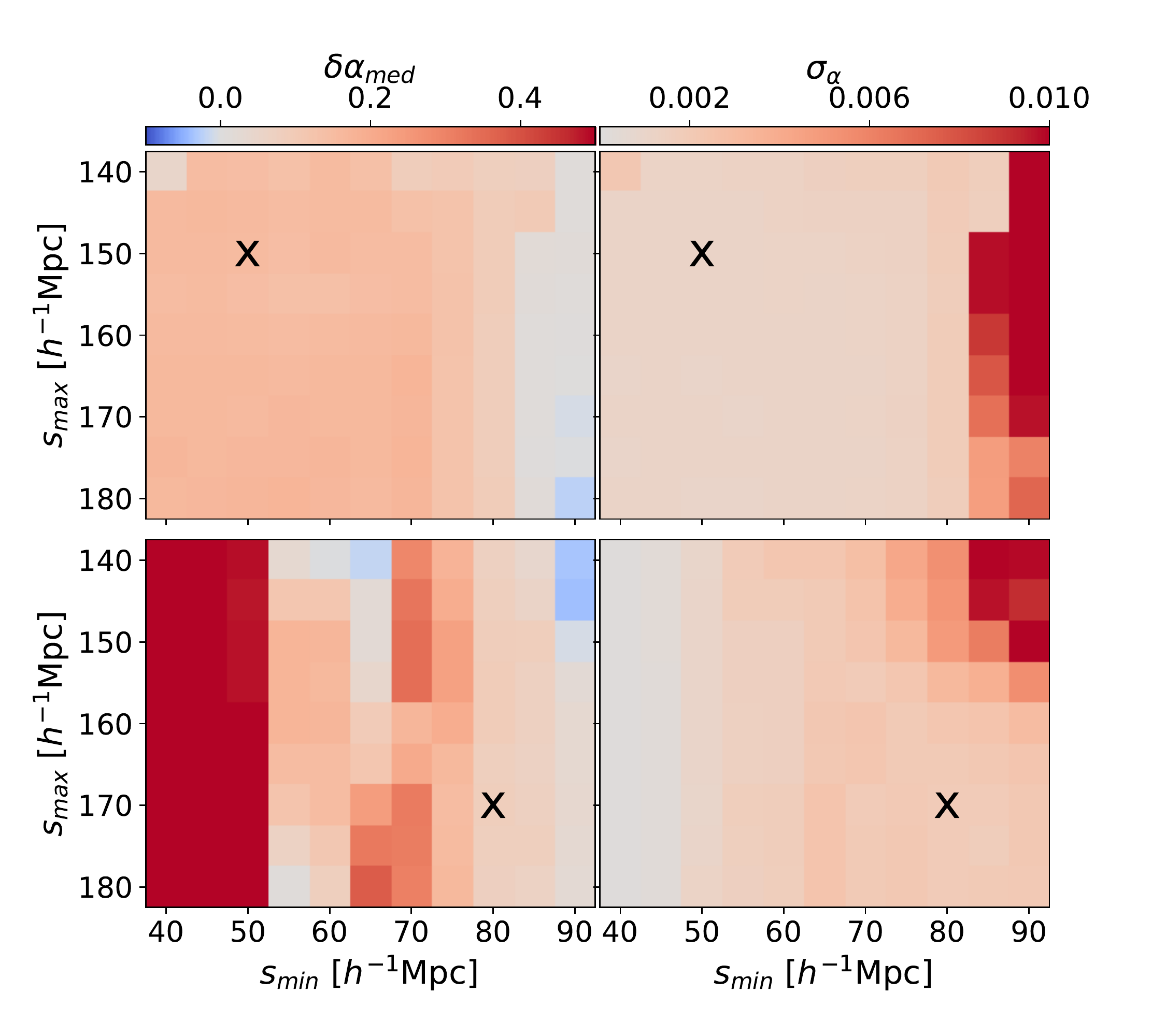}
\caption{\label{fig:EZrange} Bias $\delta\alpha_{\rm med}$ of the median of the fits from the fiducial value on the left column and evaluated 1 sigma error $\sigma_{\alpha}$ from the posterior distribution on the right, for fits with different fitting ranges. On the top results for the QSOs 2PCF of the mean of the EZmocks and on the right results for the XCF of QSOs and voids for the mean of the EZmocks. Black crosses indicate the chosen range. }
\end{figure}

Results are shown in Figure~\ref{fig:EZrange}. Minimum separation $s_{\rm min}$ of the fit varies from 40 to 90 $h^{-1}$Mpc every 5$h^{-1}$Mpc and maximum separation $s_{\rm max}$ from 140 to 180 $h^{-1}$Mpc. Following \citet{Zhao:2022aa}, we define the bias to the fiducial value $\alpha_{\rm fid}$ of the fit for the AP parameter $\alpha$ as a function of the median $\alpha_{\rm med}$ and the 1 sigma $\sigma_{\alpha}$ values of the fit posterior:
\begin{equation}
    \delta\alpha_{\rm med} = \frac{\alpha_{\rm med}-\alpha_{\rm fid}}{\sqrt{1000}\sigma_{\alpha}}.
\end{equation}
Fits for the QSO 2PCF are stable for a wide range of possibilities. We chose for consistency to adopt the range used in previous DR16 eBOSS analysis of \citet{Hou:2020aa}, a fitting range for auto-correlation of QSOs within [50,150] $h^{-1}$Mpc.

For the cross-correlation of voids and QSOs, the possible fitting ranges are more limited. Indeed usual minimum range $s_{\rm min}$ and lower are strongly affected by the exclusion effects. So to avoid contamination, we chose a conservative range of [80,170] $h^{-1}$Mpc for the XCF, where the bias and errors are reasonable when varying the minimum and maximum fitting limits by 5$h^{-1}$Mpc.

For our fiducial range, results for the mean of the EZmocks are in Table~\ref{tab:EZmean}. We also quote the maximum bias from the fitted $\alpha_{\rm med}$ when varying $s_{\rm min}$ or $s_{\rm max}$ by 5$h^{-1}$Mpc. Results are not too sensitive to a small change in the fitted range.

\begin{table}
    \centering
    \begin{tabular}{lcccc}
    \hline
         & $\alpha_{\rm med}$ & $\alpha_{\rm med}-\alpha_{\rm fid}$ & $\sigma_\alpha$ & max$|\Delta_{s}\alpha_{\rm med}|$ \\
    \hline
        $\xi_{\rm q}$ &  $1.0066^{+0.0365}_{-0.0361}$ & 0.0056 & 0.0011 & 0.0003 \\
        $\xi_{\times}$ & $1.0061^{+0.0594}_{-0.0602}$ & 0.0051 & 0.0019 & 0.0049 \\
        $\xi_{\rm mt}$ & $1.0063^{+0.0348}_{-0.0352}$ & 0.0053 & 0.0011 & 0.0011 \\
    \hline
    \end{tabular}
    \caption{ Fitting results of the AP parameter $\alpha$ on the mean of EZmocks for $R_{\min}=36$ $h^{-1}$Mpc the fiducial separation range: [50,150] $h^{-1}$Mpc, [80,170] $h^{-1}$Mpc for 2PCF $\xi_{\rm qso}$ and XCF $\xi_{\rm v}$, respectively. The multi-tracer results is noted $\xi_{\rm qso}$x$\xi_{\rm v}$. From left to right the columns are the median of the posterior with 1-sigma errors rescaled by $\sqrt{1000}$, the bias of the median of the fit to the fiducial value, the 1-sigma error of the distribution, the maximum bias from the fitted median when varying $s_{\rm min}$ or $s_{\rm max}$ by 5$h^{-1}$Mpc. }
    \label{tab:EZmean}
\end{table}

\subsection{Prior choice}
We now investigate different priors on $B$ and $\Sigma_{\rm nl}$ by fitting the EZmocks individually with the fiducial fitting range. Indeed without tighter priors, the dispersion of the errors on $\alpha$ is quite large, and there is a significant bias on average. Moreover, their dispersion is not consistent with a normal distribution as in \citet{Vargas-Magana:2013aa}.

\begin{table*}
	\centering
	\begin{tabular}{lccccccccc}
	%\begin{tabular}{|l@{\hspace{2cm}}|c@{\qquad}|c@{\qquad}|c@{\qquad}|}
		\hline 
		& $B$ Priors & $\Sigma_{\rm nl}$ Priors & $\langle\alpha_i\rangle$ & $\sigma_{\alpha_i}$ & $\langle\sigma_{\alpha, i}\rangle$ & $\frac{\sigma_{\alpha_i}-\langle\sigma_{\alpha, i}\rangle}{\sigma_{\alpha_i}}$ & $\langle g(\alpha_i)\rangle$ & $\sigma(g(\alpha_i))$ & $\langle\chi^2\rangle/$d.o.f.\\
		\hline 
\multirow{11}{*}{$\xi_{\rm q}$} & - & - & 1.023 & 0.034 & 0.111 & -2.241 & 0.037 & 0.400 & 0.984 \\
& $\pm${\it50}$\sigma$ & $\pm${\it10}$\sigma$ & {\it1.007} & {\it0.041} & {\it0.044} & {\it-0.068} & {\it-0.002} & {\it0.960} & {\it1.015} \\
& $\pm10\sigma$ & $\pm10\sigma$ & 1.007 & 0.043 & 0.039 & 0.086 & -0.012 & 1.031 & 1.038 \\
& $\pm10\sigma$ & $\pm5\sigma$ & 1.007 & 0.043 & 0.039 & 0.098 & -0.019 & 1.043 & 1.052 \\
& $\pm5\sigma$ & $\pm5\sigma$ & 1.007 & 0.043 & 0.039 & 0.101 & -0.013 & 1.042 & 1.066 \\
%& $\pm10\sigma$ & $0\rightarrow +10\sigma$ & 1.008 & 0.043 & 0.037 & 0.143 & -0.011 & 1.104 & 1.027 \\
%& $\pm50\sigma$ & $0\rightarrow +10\sigma$ & 1.007 & 0.042 & 0.043 & -0.024 & -0.021 & 1.003 & 1.004 \\
%& $\pm10\sigma$ & $0\rightarrow +5\sigma$ & 1.008 & 0.043 & 0.036 & 0.177 & -0.022 & 1.146 & 1.031 \\
%& $\pm5\sigma$ & $0\rightarrow +5\sigma$ & 1.008 & 0.043 & 0.035 & 0.185 & -0.019 & 1.150 & 1.044 \\
& - & 6.7 & 1.006 & 0.038 & 0.052 & -0.352 & 0.001 & 0.874 & 0.975 \\
& $\pm50\sigma$ & 6.7 & 1.007 & 0.042 & 0.045 & -0.072 & -0.011 & 0.970 & 0.975 \\
& $\pm10\sigma$ & 6.7 & 1.007 & 0.043 & 0.039 & 0.097 & -0.022 & 1.053 & 0.999 \\
& $\pm${\bf 5}$\sigma$ & {\bf 6.7} & {\bf 1.007} & {\bf0.044}  & {\bf 0.039}  & {\bf0.102} & {\bf-0.026} & {\bf1.058} & {\bf1.014} \\
%& $0+10\sigma$ & 6.7 & 1.002 & 0.038 & 0.056 & -0.47 & -0.041 & 0.766 & 0.992 \\
%& $0+50\sigma$ & 6.7 & 1.006 & 0.039 & 0.052 & -0.346 & -0.002 & 0.872 & 0.975 \\
& - & 5.2 & 1.006 & 0.038 & 0.051 & -0.323 & -0.017 & 0.894 & 0.974 \\
& $\pm10\sigma$ & 5.2 & 1.008 & 0.044 & 0.036 & 0.165 & -0.026 & 1.137 & 0.994 \\
	    \hline
\multirow{10}{*}{$\xi_{\times}$} & - & - & 1.020 & 0.040 & 0.100 & -1.516 & 0.055 & 0.950 & 0.871 \\
& $\pm50\sigma$ & $\pm10\sigma$ & 1.008 & 0.047 & 0.073 & -0.554 & 0.003 & 0.859 & 1.016 \\
& $\pm10\sigma$ & $\pm10\sigma$ & 1.008 & 0.049 & 0.062 & -0.253 & 0.006 & 0.925 & 1.026 \\
& $\pm10\sigma$ & $\pm5\sigma$ & 1.007 & 0.051 & 0.061 & -0.202 & 0.003 & 0.940 & 1.045 \\
& $\pm5\sigma$ & $\pm5\sigma$ & 1.008 & 0.051 & 0.060 & -0.178 & -0.004 & 0.939 & 1.057 \\
%& $\pm10\sigma$ & $0\rightarrow +10\sigma$ & 1.007 & 0.051 & 0.059 & -0.173 & 0.010 & 1.047 & 1.003 \\
%& $\pm50\sigma$ & $0\rightarrow +10\sigma$ & 1.007 & 0.045 & 0.071 & -0.578 & 0.010 & 0.871 & 0.99 \\
%& $\pm10\sigma$ & $0\rightarrow +5\sigma$ & 1.005 & 0.053 & 0.056 & -0.060 & 0.034 & 1.141 & 1.004 \\
%& $\pm5\sigma$ & $0\rightarrow +5\sigma$ & 1.004 & 0.055 & 0.055 & -0.002 & 0.036 & 1.204 & 1.022 \\
& - & 12.9 & 1.007 & 0.045 & 0.072 & -0.581 & -0.003 & 0.858 & 0.980 \\
& $\pm50\sigma$ & 12.9 & 1.006 & 0.046 & 0.072 & -0.571 & 0.003 & 0.868 & 0.980 \\
& $\pm10\sigma$ & 12.9 & 1.007 & 0.051 & 0.061 & -0.200 & -0.007 & 0.953 & 0.986 \\
& $\pm5\sigma$ & 12.9 & 1.007 & 0.051 & 0.060 & -0.181 & -0.011 & 0.950 & 1.005 \\
%& $G(B,3\sigma)$ & 12.9 & 1.008 & 0.051 & 0.06 & -0.177 & -0.004 & 0.949 & 1.035 \\
& $\pm${\bf3}$\sigma$ & {\bf12.9} & {\bf1.007} & {\bf0.051} & {\bf0.060} & {\bf-0.176} & {\bf-0.006} & {\bf0.948} & {\bf1.019} \\
        \hline
\multirow{5}{*}{$\xi_{\rm mt}$} & $\pm${\it50}$\sigma$ & $\pm${\it10}$\sigma$ & {\it1.007} & {\it0.042} & {\it0.043} & {\it-0.014} & {\it-0.020} & {\it1.043} & {\it0.896} \\
& $\pm5\sigma$ & $\pm5\sigma$ & 1.009 & 0.040 & 0.037 & 0.083 & -0.072 & 1.036 & 0.937 \\
%& $\pm50\sigma$, $\pm10\sigma$ & $0\rightarrow +10\sigma$, $0\rightarrow +5\sigma$ & 1.008 & 0.043 & 0.038 & 0.118 & -0.066 & 1.200 & 0.887 \\
& $\pm10\sigma$ & 6.7, 12.9 & 1.008 & 0.040 & 0.037 & 0.070 & -0.036 & 1.013 & 0.882 \\
& $\pm5\sigma$ & 6.7, 12.9 & 1.008 & 0.039 & 0.037 & 0.067 & -0.048 & 1.002 & 0.897 \\
& $\pm${\bf5}$\sigma$, $\pm${\bf3}$\sigma$ & {\bf6.7}, {\bf12.9} & {\bf1.009} & {\bf0.039} & {\bf0.037} & {\bf0.066} & {\bf-0.057} & {\bf1.000} & {\bf0.903} \\
        \hline
	\end{tabular} 
	\caption{Fits on correlation functions of the 1000 individual EZmocks with different parameter priors. Results are rescaled according to Equations~\ref{eq:m1} and~\ref{eq:m2}. Columns from left to right: $B$ priors, $\Sigma_{\rm nl}$ priors, the median of the individual $\alpha_i$ values, the standard deviation of the individual $\alpha_i$ fit values, the median of the individual 1-sigma errors on $\alpha_i$, the relative difference to the mean of the individual errors $\sigma_{\alpha,i}$, mean of the individual $g(\alpha_i)$ of Equation~\ref{eq:pullquantity}, the standard deviation of the individual $g(\alpha_i)$, mean reduced chi-squared of the individual fits.}
	\label{tab:indivualEZ}
\end{table*}

We then test different prior sets to find the optimal choice on our respective fiducial fitting ranges. AP parameter $\alpha$ is kept with wide flat priors. For the bias parameters $B$, we leave flat priors, but we narrow down the boundaries to $N$ times $\sigma$ around the median value given by the fit on the mean of the EZmocks for 2PCF and XCF separately, where $\sigma$ is the 1-sigma dispersion of the posterior on this parameter for the mean of the EZmocks\footnote{Fit on the mean of the EZmocks on the fiducial fitting range gives: ($B_{\rm qso}=1.336\pm0.013$,$\Sigma_{\rm nl,qso}=6.666\pm0.252$) for a fit on QSOs 2PCF, and $B_{\rm v}=8.949\pm0.242$,$\Sigma_{\rm nl,v}=12.870\pm0.588)$ for a fit on XCF.}. We also test the same kind of narrower priors on $\Sigma_{\rm nl}$ parameters. Moreover, similarly to what is done in other BAO studies \citep{Xu:2012aa,Alam:2017aa}, we fix $\Sigma_{\rm nl}$ to the median posterior value from the EZmocks mean when fitting individual EZmocks ($\Sigma_{\rm nl}=6.7$ for 2PCF and $\Sigma_{\rm nl}=12.9$ for XCF). When fitting data 2PCF, we will use the median posterior value from N-body mocks ($\Sigma_{\rm nl}=5.2$) as the BAO peak of approximated mocks as EZmocks is overdamped. It thus results in an overestimated value of $\Sigma_{\rm nl}$ in the EZmocks.

Different $\alpha$ measurements with various priors ranges are presented in Table~\ref{tab:indivualEZ} for 2PCF and XCF. As the errors for the voids are quite large, we go down to $N$=3 for XCF on the $B$ parameter.

We then chose the optimal priors from the average goodness of fit rescaled by the degree of freedom, $\langle\chi^2\rangle/$d.o.f., and the pull quantity \citep{Bautista:2020aa,Zhao:2022aa}:
\begin{equation}
    g(\alpha_i)=\frac{\alpha_i-\langle\alpha_i\rangle}{\sigma_{\alpha,i}},
\label{eq:pullquantity}
\end{equation}
where $\alpha_i$ is the median value from the posterior distribution of $\alpha$ for the $i$th EZmock realization and $\sigma_{\alpha,i}$ is its error, $\langle\alpha_i\rangle$ is the average $\alpha$ value over all EZmocks. This quantity allows us to test for the gaussianity of the results. We want to have a distribution of the $\alpha$ on the individual mocks similar to a standard distribution, i.e. a mean of 0 and a deviation of 1. 

The selected priors are in bold in the table: we chose to fix the $\Sigma_{\rm nl}$ and have narrow constraints on $B_{\rm qso}$ with $N=5$ and $N=3$ for $B_{\rm v}$. While the gaussianity of the pull quantity prefers slightly flat priors for $\Sigma_{\rm nl}$ in the 2PCF case, the reduced chi-square favours a fixed value. So for consistency with the previous analysis and with the XCF, we take fixed $\Sigma_{\rm nl}$. We note that, except in the completely free case, all results are consistent with each other. The $\alpha$ measurements are not very sensitive to the priors choices.

For the multi-tracer case, we use results from fits from separated correlations to fix $\Sigma_{\rm nl}$, and we test only a few relevant cases.

\subsection{Systematic error budget}

We refer to mocks to make a systematic error budget summarized in Table~\ref{tab:sysbudget}. A systematic bias arises from the BAO model itself. For this, we take the deviation to AP parameter true value from the EZmocks mean of our fiducial separation range of Table~\ref{tab:EZmean}. Indeed mean best-fit values from all individual N-body mocks give: $\alpha_{\rm N-body}=1.0011\pm0.0193$. The bias error is, therefore, smaller than the one from EZmocks for 2PCF. This is why we chose to quote the deviation from EZmocks for the auto-correlation alone to be conservative and consistent with the rest of the analysis with voids.

We quote a systematic bias for the maximum variation of $\alpha_{\rm med}$ when varying the fitting range of 5 $h^{-1}$Mpc. We take the value in Table~\ref{tab:EZmean} for the mean of the EZmocks.

The last systematic taken into account in the final budget is the maximum variation of the mean of the individual value of the fit on the 1000 EZmock realizations when changing the priors on $B$ and $\Sigma_{\rm nl}$. We take a conservative choice and take as a reference for the systematic largest flat priors indicated in italic in Table~\ref{tab:indivualEZ}.

The three contributions are added in quadrature to obtain the final systematic error $\sigma_{\rm syst}$.

\begin{table}
    \centering
    \begin{tabular}{lcccc}
    \hline
         & $\alpha_{\rm fit} - \alpha_{\rm fid}$ & max$|\Delta_{s}\alpha_{\rm med}|$ & max$|\Delta_{\rm prior}\langle\alpha_{i}\rangle|$ & $\sigma_{\rm syst}$\\
    \hline
        $\xi_{\rm q}$ & 0.0056 & 0.0003 & 0.0001 & 0.0056 \\
        $\xi_{\rm mt}$ & 0.0053 & 0.0011 & 0.0009 & 0.0055\\
    \hline
    \end{tabular}
    \caption{Systematic error budget. Different columns are the different constributions to the total error $\sigma_{\rm syst}$ for QSO 2PCF and the multi-tracer analysis.}
    \label{tab:sysbudget}
\end{table}

\subsection{Change in radius cut}
We test the template used for the BAO model and analysis robustness by observing the changes induced by a small variation of the minimum radius cut of the voids. For this, we use the same template model as for the fiducial analysis with $R_{\rm min}=36$ $h^{-1}$Mpc and vary $R_{\rm min}$ of the EZmocks XCF by 2 $h^{-1}$Mpc. 

Table~\ref{tab:EZradiuscut} gives the results for the mean of the EZmocks for $R_{\rm min}=34$ $h^{-1}$Mpc and $R_{\rm min}=38$ $h^{-1}$Mpc. As mentioned, the template is not adapted for those radius cuts, so it inserts an expected mild bias compared to the fiducial measurements of Table~\ref{tab:EZmean} for the XCF. For the multi-tracer approach with XCF and 2PCF, the bias is small: a small change in the radius cut inserts, therefore a reasonable bias.

\begin{table}
    \centering
    \begin{tabular}{lccc}
    \hline
         & $\alpha_{\rm med}$ & $\sigma_\alpha$ & $\alpha_{\rm med}-\alpha_{\rm 36}$ \\
    \hline
        $\xi_{\times}$, $R_{\rm v}>34$ &  $1.0164^{+0.0545}_{-0.0589}$ & 0.0018 & 0.0103 \\
        $\xi_{\times}$, $R_{\rm v}>38$ & $0.9960^{+0.0615}_{-0.0639}$ & 0.0020 & -0.0101 \\
        $\xi_{\rm mt}$, $R_{\rm v}>34$ & $1.0083^{+0.0360}_{-0.0347}$ & 0.0011 & 0.0020 \\
        $\xi_{\rm mt}$, $R_{\rm v}>38$ & $1.0052^{+0.0353}_{-0.0367}$ & 0.0011 & -0.0011\\
    \hline
    \end{tabular}
    \caption{Fitting results of the AP parameter $\alpha$ on the mean of EZmocks for the fiducial separation range with two different minimum voids radius cut for the XCF. From left to right, the columns are the median of the posterior with 1-sigma errors rescaled by $\sqrt{1000}$, the bias of the median of the fit to the fiducial value, the 1-sigma error of the distribution, the bias of the median of the fit to the value for the fiducial cut of $36$ $h^{-1}$Mpc.}
    \label{tab:EZradiuscut}
\end{table}

\subsection{Results on EZmocks}
Let us now compare the BAO results of the QSOs auto-correlation and the multi-tracer joint fit of the 2PCF and the XCF. We consider the individual 1000 EZmocks realisations in the fiducial case (minimum radius cut, separation range and priors), i.e. the bold lines in Table~\ref{tab:indivualEZ}.

\begin{figure}
\centering
\includegraphics[width=0.8\columnwidth]{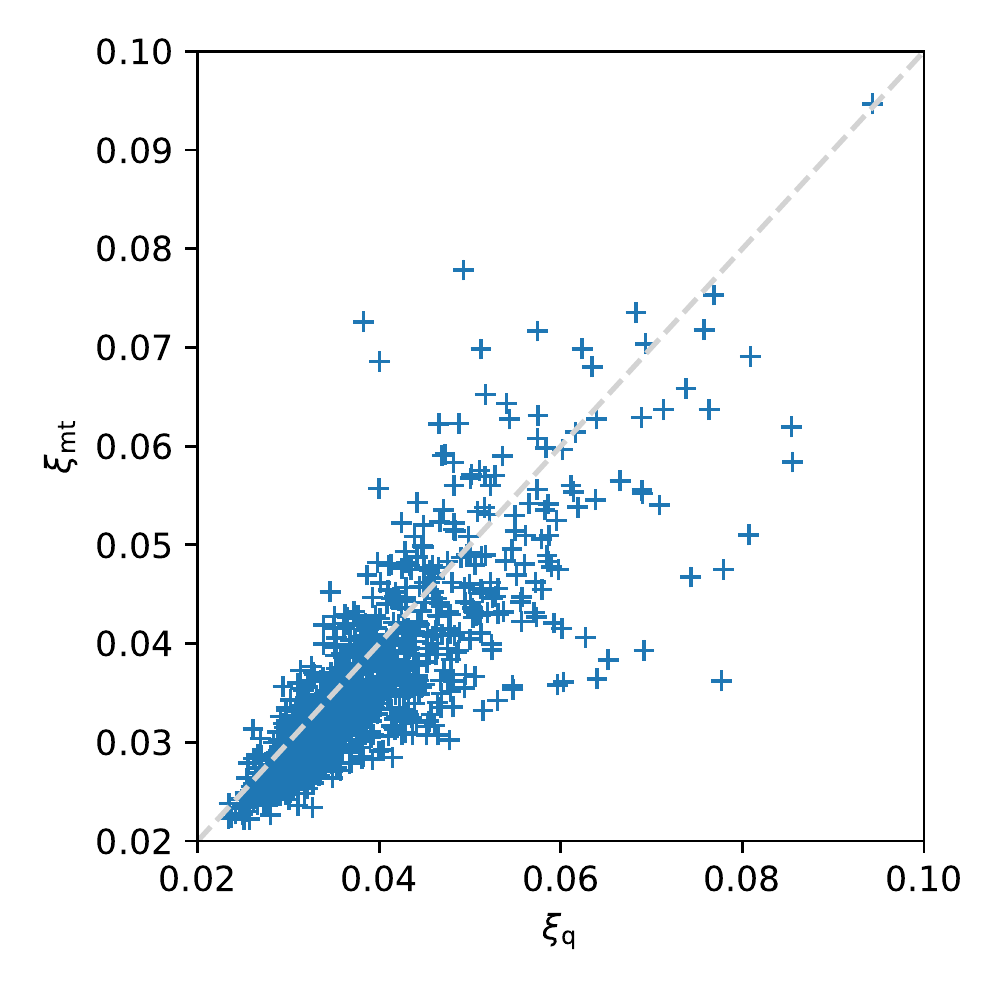}
\caption{\label{fig:EZres} Errors from the 2PCF fits against the multi-tracer results which fits jointly the 2PCF and XCF. }
\end{figure}

We define the relative difference in errors between the two analyses:
\begin{equation}
    \delta_i=\frac{\sigma_{\alpha,i,\rm q}-\sigma_{\alpha,i,\rm mt}}{\sigma_{\alpha,i,\rm q}}
\label{eq:relErr}
\end{equation}
where $\sigma_{\alpha,i,\rm q}$ is the 1-sigma distribution error on $\alpha_i$ for the 2PCF case, and $\sigma_{\alpha,i,\rm mt}$ in the multi-tracer case. This statistic is presented in Table~\ref{tab:EZdelta} for the individual EZmocks. Figure~\ref{fig:EZres} compares the errors from fits of QSO 2PCF only and those from the multi-tracer version.

There is an average of about 5$\%$ improvement with the contribution of voids in the analysis. A smaller error for the multi-tracer case is observed for around 70$\%$ of the EZmocks realisations. Taking only the improved mocks gives, on average better errors of 11.22$\%$. Fitting QSO voids jointly with QSOs allows, therefore, a small improvement for most of the EZmocks on the same sample of data.

In the previous study of \citet{Zhao:2022aa} for eBOSS ELG and LRG samples, the best results on EZmocks were reported to give a larger average improvement ($\sim8\%$). However, we note that in this case, the void auto-correlation was also jointly fitted and helped reduce the uncertainties. Closer statistics are found when comparing the joint fit with the cross-correlation only. Moreover, with QSOs, some exclusion effects might still play an important role, and this makes the extraction of the BAO information more difficult.

\begin{table}
    \centering
    \begin{tabular}{lccc}
    \hline
         & $\langle\delta_i\rangle$ & ($\langle\delta_i|\delta_i>0\rangle$) & $\#(\delta_i|\delta_i>0)$ \\
    \hline
        $\xi_{\rm mt}$ & 5.41$\%$ & (11.22$\%$) & 71.6$\%$ \\
    \hline
    \end{tabular}
    \caption{Mean relative difference $\delta_i$ of Equation~\ref{eq:relErr} for the individual realisations of EZmocks, mean relative difference when $\delta_i$ is positive, and proportion of realisations for which $\delta_i$ is positive. }
    \label{tab:EZdelta}
\end{table}

\section{Results}\label{sec:results}

In this Section, we present the results of the eBOSS DR16 QSO data sample. Table~\ref{tab:datares} displays the $\alpha$ measurement and its derived value for our input cosmology, the volume-averaged distance of Equation~\ref{eq:DV}. Fits are made on the fiducial fitting range with the selected priors for $B$ and $\Sigma_{\rm nl}$. For QSO 2PCF data fit, we fix $\Sigma_{\rm nl}$ to the value given by N-body mocks. Voids are selected according to their radius with a hard minimum cut range; see Equation~\ref{eq:Rmin}.

\subsection{eBOSS DR16 QSO sample}
\begin{figure}
\centering
\includegraphics[width=1.0\columnwidth]{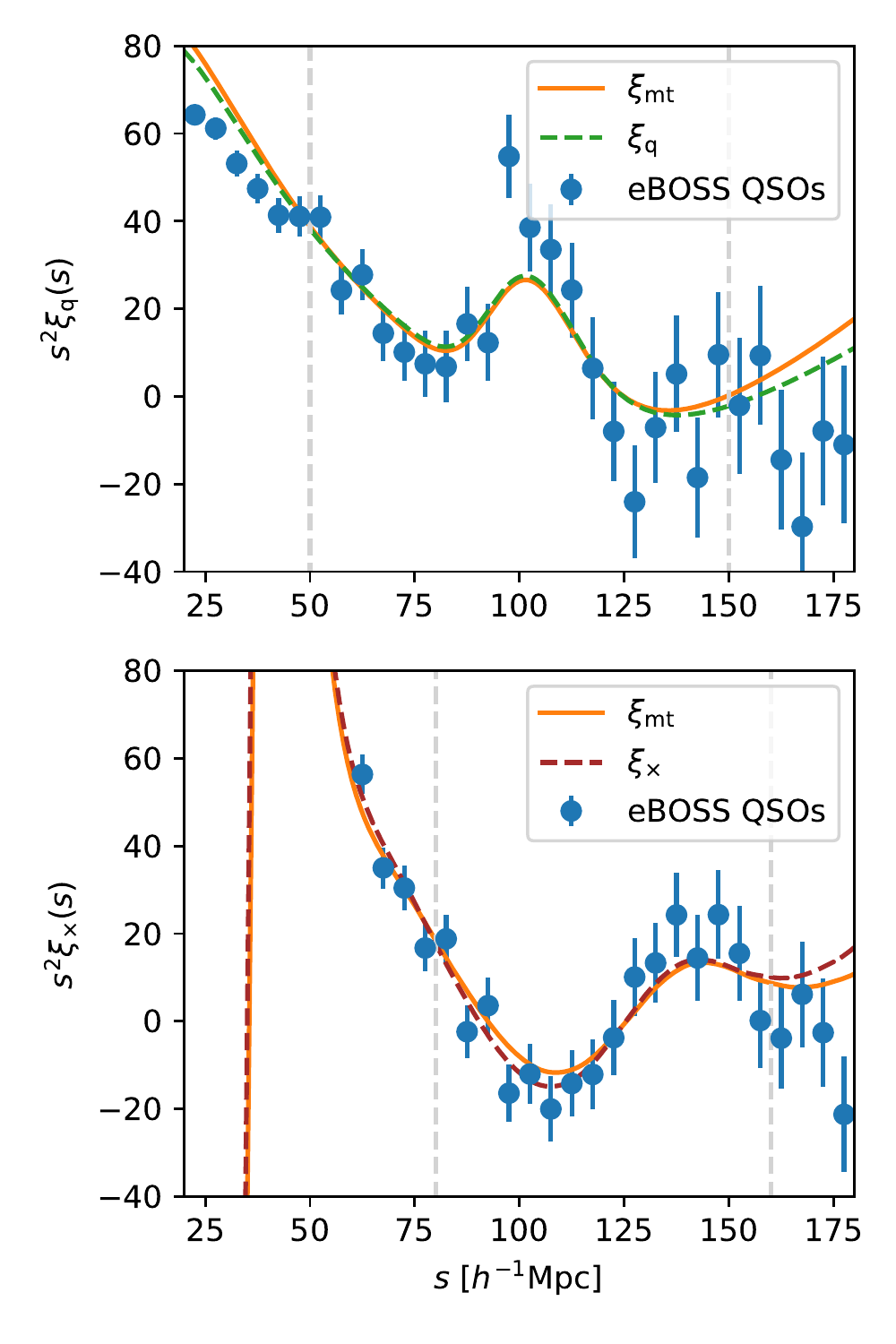}
\caption{\label{fig:modelData} Best-fit models as fitted for the 2PCF or XCF alone or jointly. }
\end{figure}

For data, we observe very similar results from QSOs only or adding voids. The reduced chi-squared is slightly better for the multi-tracer case. However, errors are not improved by the 2PCF joint fit with XCF compared to 2PCF alone. We note, moreover, that $\Sigma_{\rm nl, \times}$ was estimated from EZmocks that tend to overestimate it. A better determination of $\Sigma_{\rm nl, \times}$ could lead to better results. The best-fitting BAO models are shown in Figure~\ref{fig:modelData}. The data are well fitted on the fitting range in all cases. Results are consistent with the isotropic measurement of \citet{Neveux:2020aa} on the same QSO eBOSS sample, in particular we recover similar errors \citep[see also][]{Hou:2020aa}.

EZmocks results suggest that data measurement lies in the 30$\%$ hazard without improvement observed with a joint fit with the cross-correlation of voids. To recreate the randomness of the sampling of data, we create 25 subsamples of the eBOSS QSOs by removing 1/25 of the area with equal numbers of QSOs different for each of the samples. We then fit them in the same way as for the total sample. 

Table~\ref{tab:SplittedData} gathers the measurements for the 25 data subsamples. The average value is consistent with the data alone. Moreover, we have an average improvement of about 2$\%$ for almost 70$\%$ of the realisations. This result is in total agreement with the EZmocks. It implies that voids could still bring a small improvement for future QSOs surveys. Indeed an improvement is expected, but for a specific data sample, the improvement is not necessarily seen due to cosmic variance.

\begin{table}
    \centering
    \begin{tabular}{lcccc}
    \hline
         & $\alpha_{\rm fit}$ & $\sigma_{\rm syst}$ & $\chi^2/$d.o.f. &  $D_V(z=1.48)/r_s$\\
    \hline
        $\xi_{\rm q}$ & $1.0172_{-0.0201}^{+0.0207}$ & 0.0056 & 1.49 & $26.298\pm0.547$ \\
        $\xi_{\rm mt}$ & $1.0171_{-0.0196}^{+0.0212}$ & 0.0055 & 1.16 & $26.297\pm0.547$ \\
    \hline
    \end{tabular}
    \caption{Results on the eBOSS QSO data sample for the standard 2PCF analysis and with the void contribution multi-tracer with XCF. Median of the posterior of the fitted $\alpha$ parameter and the 16th and 84th percentiles. Total systematic error. The goodness of fit is rescaled by the degree of freedom. The volume-averaged distance at the effective redshift.}
    \label{tab:datares}
\end{table}

\begin{table}
    \centering
    \begin{tabular}{lccc}
    \hline
         & $\langle\alpha_i\rangle$ & $\langle\delta_i\rangle$ &  $\#(\delta_i|\delta_i>0)$ \\
    \hline
        $\xi_{\rm mt}$ & 1.016$\pm$0.021 & 2.09$\%$ & 68.0$\%$ \\
    \hline
    \end{tabular}
    \caption{Mean $\alpha$ measurement and 1-sigma dispersion for the 25 subsampled data, relative difference $\delta_i$ of Equation~\ref{eq:relErr} and proportion of realisations for which $\delta_i$ is positive. }
    \label{tab:SplittedData}
\end{table}

\subsection{DESI-like volume survey forecasts}
We further provide a forecast for a QSO survey with a similar effective volume to that of DESI for BAO constraints from QSOs. We repeat the same BAO analysis on 1000 EZbox with BAO.

The effective volume of EZbox is very close to the Year 5 DESI effective volume for an area of 14'000 deg$^2$ \citep{DESI-Collaboration:2016aa} of QSOs. Therefore we directly use the covariance made from the 1000 EZbox without rescaling. 

We perform BAO measurements on the 1000 individual realisations for the QSOs 2PCF alone and jointly fitted with their cross-correlation. Following the results of the SNR test of section~\ref{EZboxR} for the EZbox, the void radius cut is chosen to be 31 $h^{-1}$Mpc. For the BAO model, we recreate an appropriate template. The clustering of the boxes is consistent with that of the light-cone mocks and the data. In this case, it is appropriate to use the Lagrangian mocks generated for the light-cone mocks, but without radial selection and survey geometry cut, i.e. in their boxes format. The cross-power spectra are then computed for the optimal minimum radius cut of 31 $h^{-1}$Mpc. Measurements are gathered in Table~\ref{tab:EZboxdelta}.

We recover the same results as for the EZmocks. About 68$\%$ of the EZbox realisations have an error reduction when fitting the 2PCF and XCF simultaneously. This improvement is 4.9$\%$ on average. This means that increasing the volume, i.e. decreasing the statistical errors, does not help to have a general improvement of the BAO error by adding voids. This might be due to the low density of the QSOs samples. Therefore we expect the results from actual DESI data to be better, as the density of the QSO boxes is still lower than the expected QSO density of DESI.

\begin{table}
    \centering
    \begin{tabular}{lccccc}
    \hline
         & $\langle\alpha_i\rangle$ & $\sigma_{\alpha_i}$ & $\langle\sigma_{\alpha, i}\rangle$ & $\langle\delta_i\rangle$ & $\#(\delta_i|\delta_i>0)$ \\
    \hline
        $\xi_{\rm mt}$ & 1.003 & 0.008 & 0.008 & 4.90$\%$ & 68.2$\%$ \\
    \hline
    \end{tabular}
    \caption{Multi-tracer fitting results for the 1000 individual realisations of EZbox with BAO. Columns from left to right: median of the individual AP parameter $\alpha_i$ fitting values, the standard deviation of the individual $\alpha_i$ fit values, the median of the individual 1-sigma errors on $\alpha_i$, mean relative difference $\delta_i$ of Equation~\ref{eq:relErr} and proportion of realisations for which $\delta_i$ is positive. $\Sigma_{\rm nl}$ of the fits are fixed from the EZbox mean fits, and we use a $\pm$10$\sigma$ priors on $B$.}
    \label{tab:EZboxdelta}
\end{table}

\section{Conclusions}\label{sec:conclusions}
In this paper, we proposed a void analysis of the QSO eBOSS DR16 sample with voids. Due to the low density of the sample, the minimum size of the void required to mitigate the contamination by voids-in-clouds is about twice the size for the previous analysis \citep{Zhao:2020aa,Zhao:2022aa} with the same void definition.

To understand the BAO signal from the void correlations, we produced EZmocks with and without BAO signature. This allowed us to choose the optimal radius cut to increase the BAO signal and minimize the noise. We are able to observe a negative BAO peak in the cross-correlation of QSOs and voids. However, we did not detect any signal in the auto-correlation of voids as  geometric exclusion effects affect the BAO scale, since we are considering very large voids. We note that we explored other ways of extending the void catalogue including voids with smaller radii based on QSO local density arguments to increase the number density and alleviate the void exclusion effects. However, some biases appeared in this process, which make such attempts still unreliable. We leave a further investigation on this for future work.

We presented a multi-tracer fit of the 2PCF and XCF jointly. For EZmocks, the errors decreased for 70$\%$ of the realisations when voids were jointly fit with QSOs. We report an average of around 5$\%$ error improvement for the EZmocks. While we found less improvement than for the other tracers as LRGs and ELGs by adding the contribution of voids \citep{Zhao:2022aa}, we argued that it might be caused by the difficulty of extracting the BAO information due to remaining void exclusion effects. Moreover, the auto-correlation of voids that have a non-negligible constraining power was not included.  

For eBOSS QSOs sample data, no improvement was measured including voids. Our analysis showed the same behaviour as for EZmocks when we downsample the data into 25 subsamples. This confirmed that the result for the data is caused by cosmic variance.

We finally presented a forecast for 
the next batch of surveys like DESI, which will release a large sample of QSOs \citep{DESI-Collaboration:2016aa, DESI-Collaboration:2016ab}. Our results demonstrate that voids can still improve the isotropic BAO AP parameter for those data by almost 5$\%$, a result which remains stable even if the volume is increased. Better improvement is expected for future QSO surveys with a higher number density such as J-PAS \citep{Benitez:2014aa} or WEAVE \citep{Dalton:2016aa,Pieri:2016aa}.
Hence, we conclude, that voids can be potentially useful to further increase the BAO detection from forthcoming QSO catalogues.

\section*{Acknowledgements}
AT, CZ, DFS and AV acknowledge support from the SNF grant 200020$\_$175751. %??? TO DO 207379 !!!

Funding for the Sloan Digital Sky Survey IV has been provided by the Alfred P. Sloan Foundation, the U.S. Department of Energy Office of Science, and the Participating Institutions. SDSS-IV acknowledges support and resources from the Center for High- Performance Computing at the University of Utah. The SDSS web site is www.sdss.org.
SDSS-IV is managed by the Astrophysical Research Consortium for the Participating Institutions of the SDSS Collaboration including the Brazilian Participation Group, the Carnegie Institution for Science, Carnegie Mellon University, the Chilean Participation Group, the Ecole Polytechnique Federale de Lausanne (EPFL), the French Participation Group, Harvard-Smithsonian Center for Astrophysics, Instituto de Astrofisica de Canarias, The Johns Hopkins University, Kavli Institute for the Physics and Mathematics of the Universe (IPMU) University of Tokyo, the Korean Participation Group, Lawrence Berkeley National Laboratory, Leibniz Institut für Astrophysik Potsdam (AIP), Max-Planck-Institut für Astronomie (MPIA Heidelberg), Max-Planck-Institut für Astrophysik (MPA Garching), Max-Planck-Institut für Extraterrestrische Physik (MPE), National Astronomical Observatories of China, New Mexico State University, New York University, University of Notre Dame, Observatário Nacional / MCTI, The Ohio State University, Pennsylvania State University, Shanghai Astronomical Observatory, United Kingdom Participation Group, Universidad Nacional Autónoma de México, University of Arizona, University of Colorado Boulder, University of Oxford, University of Portsmouth, University of Utah, University of Virginia, University of Washing- ton, University of Wisconsin, Vanderbilt University, and Yale University.

This research used resources of the National Energy Research Scientific Computing Center, a DOE Office of Science User Facility supported by the Office of Science of the U.S. Department of Energy under Contract No. DE-AC02-05CH11231.

%%%%%%%%%%%%%%%%%%%%%%%%%%%%%%%%%%%%%%%%%%%%%%%%%%
\section*{Data Availability}

Mock catalogues and data sample used in this paper are available via the SDSS Science Archive Server. In particular the QSO eBOSS sample can be find here: \url{https://data.sdss.org/sas/dr17/eboss/qso/DR16Q/}. For EZmocks, they are in: \url{https://data.sdss.org/sas/dr17/eboss/lss/EZmocks/}. Codes used for the analysis, correlation computations and void finder are all available, as indicated in the paper's footnotes.



\bibliographystyle{mnras}
\bibliography{Ref}

%%%%%%%%%%%%%%%%%%%%%%%%%%%%%%%%%%%%%%%%%%%%%%%%%%

%%%%%%%%%%%%%%%%% APPENDICES %%%%%%%%%%%%%%%%%%%%%

%\appendix


%%%%%%%%%%%%%%%%%%%%%%%%%%%%%%%%%%%%%%%%%%%%%%%%%%

% Don't change these lines
\bsp	% typesetting comment
\label{lastpage}
\end{document}